\documentclass[draft]{agujournal2019}
\usepackage{url} 
\usepackage{lineno}
\usepackage[inline]{trackchanges} 
\usepackage{soul}
\usepackage{amsmath,stmaryrd}
\usepackage{multirow}
\draftfalse

\journalname{Journal of Advances in Modeling Earth Systems (JAMES)}

\begin{document}

%
%


\title{Identifying Three-Dimensional Radiative Patterns Associated with Early Tropical Cyclone Intensification}

%
%




\authors{Frederick Iat-Hin Tam\affil{1,2}, Tom Beucler\affil{1,2}, James H. Ruppert Jr.\affil{3}}

\affiliation{1}{Faculty of Geosciences and Environment, University of Lausanne, Switzerland}
\affiliation{2}{Expertise Center for Climate Extremes, University of Lausanne, Switzerland}
\affiliation{3}{School of Meteorology, University of Oklahoma, Norman, OK, USA}




\correspondingauthor{Frederick Iat-Hin Tam}{ft21894@gmail.com}




\begin{keypoints}
\item We extract three-dimensional radiative heating patterns critical to tropical cyclogenesis from WRF simulations

\item Our data-driven model pinpoints time periods where radiative heating has a dominant effect on early TC intensification

\item Downshear longwave anomaly patterns are more relevant to the genesis of Haiyan than azimuthally uniform ones
\end{keypoints}

%
%

%
%



\begin{abstract}
Cloud radiative feedback impacts early tropical cyclone (TC) intensification, but limitations in existing diagnostic frameworks make them unsuitable for studying asymmetric or transient radiative heating. We propose a linear Variational Encoder-Decoder (VED) framework to learn the hidden relationship between radiative anomalies and the surface intensification of realistic simulated TCs. The uncertainty of the VED model identifies periods when radiation has more importance for intensification. A close examination of the radiative pattern extracted by the VED model from a 20-member ensemble simulation on Typhoon Haiyan shows that longwave forcing from inner core deep convection and shallow clouds downshear contribute to intensification, with deep convection in the downshear-left quadrant having the most impact overall on the intensification of that TC. Our work demonstrates that machine learning can aid the discovery of thermodynamic-kinematic relationships without relying on axisymmetric or deterministic assumptions, paving the way for the objective discovery of processes leading to TC intensification in realistic conditions.
\end{abstract}

\section*{Plain Language Summary}
How clouds react to heating from the sun and Earth's surface, called Cloud Radiative Feedback (CRF), greatly affects how quickly tropical cyclones intensify in their early stage. Traditional methods to isolate CRF have limitations. In this manuscript, we demonstrate that machine learning can be applied to this problem and allows us to isolate, quantify, and visualize CRF with better spatial details than what is possible with traditional methods. One of the many things our machine-learning model can do is identify times when CRF is important and times when other processes are more important. Applying our model to two historical tropical cyclones shows that CRF in certain regions close to the tropical cyclones' centers are crucial for their early intensification. This helps us understand how these storms form in real-world conditions.

%
%

%


%
%
%
%

\section{Introduction}

Recent advances in numerical weather prediction (NWP) models brought substantial improvements to tropical cyclone (TC) track predictions ~\cite{Landsea_Cangialosi2018}. However, predicting TC intensity ~\cite{DeMaria_etal2014,Cangialosi_etal2020} remains challenging. One of the most elusive aspects of TC strength evolution relates to the formation (`genesis') and early intensification of TCs, which involves multiscale interaction between ambient large-scale circulation and small-scale organized convective clusters ~\cite{narenpitak2020role}. 

Radiative feedback has been shown to be one of the leading factors in the spontaneous self-aggregation of tropical oceanic convection in non-rotating simulations of radiative-convective equilibrium ~\cite<RCE; >{MullerBony2015, Wing_etal2017, Fan_etal2021}. Including background rotation in idealized RCE experiments results in self-aggregation that forms rotating, TC-like systems \cite{Carstens2020TropicalCF}. Recent observational and modeling studies reveal four radiative feedbacks that can potentially affect TC intensity: (1) longwave environmental destabilization ~\cite{MelhauserZhang2014}, (2) radial gradient in radiative heating between cloudy areas and cloudless surroundings ~\cite{GrayJaccobson1977, Nicholls2015, Smith_etal2020},(3) trapping of longwave radiation by clouds ~\cite{Bu_etal2014, Ruppert_etal2020}, and (4) diurnal radiative forcing at TC cloud anvil  ~\cite{MelhauserZhang2014, Ruppert_ONeill2019}. Processes (2) and (3) are linked, whereby the local greenhouse trapping by clouds causes a radiative heating gradient relative to the cloudless surroundings. Although smaller in magnitude than latent heating, radiation significantly affects TC intensity by influencing the spatial distribution of convection \cite{dai2023longwave}. During the genesis phase, removing processes (2) and (3) delay TC formation by a factor of 2 to 3 \cite{muller2018acceleration,Ruppert_etal2020,Smith_etal2020,rios2020impacts}. Processes (2) and (3) contribute to early TC intensification through radiation-driven transverse circulations \cite{Nicholls2015}. These circulations sustain TC convection by moistening the inner core \cite{Ruppert_etal2020}, enhancing mid-level circulations \cite{yang2022cloud}, and protecting against shear-related ventilation \cite{rios2020impacts}.

The effect of radiation on TC evolution is usually examined through the column Moist Static Energy (MSE) variance budget ~\cite{Wing_etal2017, Wu_etal2021} and axisymmetric balanced dynamics ~\cite{Ruppert_etal2020,navarro2017balanced}. Thermal forcing from diabatic processes drives the TC out of thermal wind balance; the secondary circulation then restored the balanced state ~\cite{Willoughby1979}. The Sawyer-Eliassen Equation ~\cite<SEQ;>{PendergrassWilloughby2009} can diagnose such secondary circulations. However, the SEQ has limitations because it requires the TC to be in a quasi-steady state. Transient thermal forcings are smoothed out to satisfy the quasi-steady state assumption, and the SEQ solutions are only available in a two-dimensional radius-height plane. The lack of azimuthal context makes the SEQ framework suboptimal to study TC genesis pathways under shear \cite{rogers2020precipitation,nam2021multiscale}, and the role of asymmetric convection in TC intensity changes \cite{rogers2013airborne,zagrodnik2014rainfall,xu2017relationships}.

While the role of asymmetric convection on TC intensification can be addressed with budget analysis \cite<e.g., >{wu2004effects}, these methods require calculating multiple terms with spatial and temporal differentiation that are not typically available in standard model outputs. It is also difficult to close these budgets with coarse-resolution model outputs. To address these challenges, we developed a machine learning (ML) framework to advance our understanding of how spatial patterns affect TC intensification in cloud-resolving model outputs \cite{mcgovern2019making}. Specifically, we identify three-dimensional radiative heating structures relevant to TC intensification in datasets lacking online budget calculations and with coarse temporal resolution. In an ML modeling framework, these structures can be found with post-hoc attribution methods. Applying Explainable Artificial intelligence (XAI) tools to ML algorithms led to the discovery of potential vorticity patterns favoring TC rapid intensification during trough-TC interaction \cite{Fischer_etal2019} and outgoing longwave radiation patterns enhancing the predictability of North Atlantic extratropical circulation \cite{MayerBarnes2021}. In contrast to the post-hoc explanation methods, recent studies \cite<e.g., >{barnes2022looks, Behrens_etal2022} have leveraged specialized architectures akin to the Variational Encoder-Decoder architecture used in this study to find hidden structures in climate model outputs without using XAI tools, alleviating attribution uncertainty \cite{mamalakis2023carefully}. 

Using complex ML models for scientific discovery poses notable challenges. First, these models are highly nonlinear and often not interpretable, which limits their trustworthiness. Second, complex ML models demand substantial data quantities for robust performance, which proves problematic when dealing with limited data samples. For example, we rely on limited model hindcasts in this work because finding the 3D radiative patterns relevant for genesis requires training on full 3D radiative heating structures - a variable not available in observations and reanalysis data. To address these challenges, we have developed a linear variational encoder-decoder (VED) model that fulfills the main traits of interpretable ML models: (i) simulatability, ensuring simplicity for comprehensibility, and (ii) decomposability, meaning that the model has meaningful inputs and parameters \cite{Lipton2018, MarcinkevičsVogt2023}. 

This research investigates the potential of interpretable ML models to discover realistic radiative heating structures relevant to TC intensification from limited data. We offer physical insights on how these structures may contribute to early TC intensification. We will also show that we can use the uncertainties and bias in the model predictions to isolate the exact period where radiative feedback is most relevant to TC intensification.

\section{Data}
\subsection{Convection-Permitting Hindcasts of Two Tropical Cyclones}
We analyze a 20-member set of WRF Version 4.3.1 \cite{skamarock2008description} ensemble hindcast simulations of Typhoon Haiyan (2013) and a small set of simulations of Hurricane Maria (2017), ran at a convective-permitting resolution (3 km). The same set of simulations for Maria in \citeA{Ruppert_etal2020} is analyzed to evaluate if ML models can learn the same physical conclusion in that paper - disabling Cloud Radiative Feedback (CRF) disrupts tropical cyclogenesis. This set of simulations includes one Control (CTRL) experiment and 4 ``no-CRF'' sensitivity simulations. ``No-CRF'' simulations are produced by running the CTRL restart files at different times, albeit setting all hydrometeor terms in the longwave and shortwave schemes to be zero. The CTRL simulation was initialized from GEFS analysis and integrated from September 14, 2017, 1200 UTC to September 20, 2017, 1200 UTC. Readers are referred to \citeA{Ruppert_etal2020} for more details regarding the setup of the Maria experiments. 

We use the Haiyan ensemble simulations to analyze the role of CRF in realistic conditions, i.e., without the no-CRF experiments. These simulations are constructed by dynamically downscaling the National Center for Environmental Prediction's Global Ensemble Forecast System (GEFS) ensemble member outputs. The primary source of variability in the ensemble comes from the interaction between convection and slight variability in the GEFS synoptic conditions. The Haiyan WRF simulations are integrated from November 1, 2013, 00 UTC to November 8, 2013, 00 UTC, with a two-nested, 15km-3km horizontal grid spacing fixed model domain. The inner domain is around 3600x2200 km in size. Radiation is treated with the Rapid Radiative Transfer Model for GCMs ~\cite<RRTMG;>{RRTMG2008}, and microphysics is treated using the Thompson and Eidhammer scheme ~\cite{Thompson2014}. Other model physics are configured identically to the Maria simulations. The model contains 55 stretched vertical levels and is topped at 10 hPa. These simulations are assigned integer labels from 0 to 19. All simulations produce outputs at an hourly interval; these outputs are post-processed into a TC-relative framework by tracking the local maxima in 700-hPa absolute vertical vorticity, spatially smoothed with a 1.5-degree boxcar filter, and temporally filtered with a 3-point Gaussian filter to remove noise.

For Haiyan, while we use all the data for training and evaluation, we keep the analysis tractable by focusing on two ensemble members: Member 2, which intensifies at a quicker rate, and Member 11, which intensifies at a slower rate.

\subsection{Cross-validation strategy}
Following best machine learning practices, we divide the data into training, validation, and test sets. Evaluating model skills on data unseen during training ensures good prediction skills for out-of-sample data. This section discusses how we perform the data splitting. 

Since ensemble simulations and sensitivity experiments can be treated as different realizations of the same physical system responding to slightly different forcing, we opt for a data-splitting strategy based on ensemble member labels (Haiyan) and experiment labels (Maria). 

For Haiyan, we use 80$\%$ of the ensemble outputs (16 experiments) for training, and 20$\%$ of the data (4 experiments) for validation and testing. Two experiments that are not strongly correlated to the other experiments are left out at first to create an independent test set; the remaining 18 experiments are partitioned into training and validation subsets by randomly generating a list of two numbers between 1 and 20; the two numbers are then used as references to separate the validation set from the training set. This data-splitting procedure is repeated forty times to create forty different sets of training data, which enables evaluations of the model variability associated with the choice of data split. Our test set is truly independent because the two test experiments are never used in the training or validation set for all data splits. 

We slightly altered the data-splitting strategy for Maria due to a lack of samples. The Control (CTRL) simulation is always included in the training set because it has the most samples and represents how the TC evolves in realistic conditions where CRF always exists. The ``NCRF-36h'' experiment is used as the test dataset amongst the four remaining sensitivity experiments because the storm intensity changes in that experiment depart most from the CTRL simulation. We randomly split the other experiments into a portion that is merged into the training set (2 experiments) and the other portion for model validation (1 experiment). The cross-validation strategy for Maria yields three different data splits to test the ability of the trained ML models to depict the counterfactual scenario of TC evolution without cloud radiative feedback (CRF).

\section{Methodology}

\subsection{Machine Learning Framework}
\begin{figure*}
    \includegraphics[width=\textwidth]{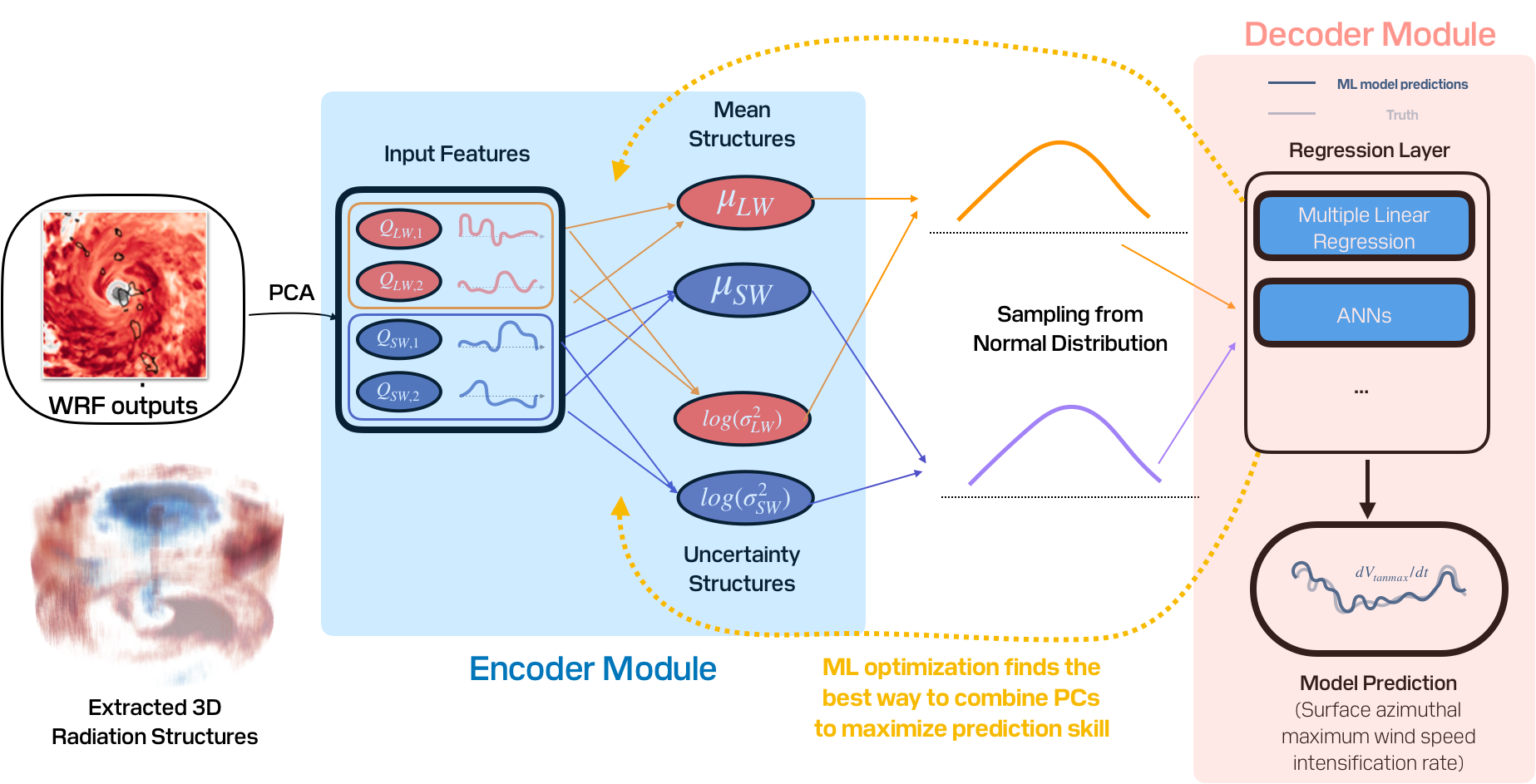}
    \caption{The interpretable linear VED framework proposed in this study combines a pattern-finding encoder and a decoder for TC intensification rate prediction. The first linear layer in the encoder modules combines different radiation structural information in the PCs into the time evolutions of the mean structures and uncertainty structures. A random sampling of the normal distributions with mean and log variances conditioned on the inputs introduces uncertainty to a decoder module that predicts the 24-hour surface wind intensification prediction. There is flexibility in the choice of the decoder architecture, from simple multiple linear regressors to a complex nonlinear Artificial Neural Network (ANN). Optimization of model weights with loss functions tries to minimize the absolute error between the truth and the predictions from the decoder based on the learned patterns.}
    \label{fig:schematic}
\end{figure*}

Our analysis leverages machine learning (ML) to identify radiative spatial features relevant to TC intensification from WRF output. We adopt an interpretable, stochastic linear VED model to discover such features. Latent heating is not used as an input because it is treated as an internal response to external forcings such as radiation in the moist static energy variance budget \cite{WingEmanuel2014,wing2022acceleration}. A schematic diagram of our framework is provided in Fig~\ref{fig:schematic}. The framework is divided into two parts: a learned encoder and a learned decoder for TC surface intensity predictions. The learned encoder distills radiative information relevant to TC intensity predictability into a limited number of structures. We constrain the number of extracted structures (one per variable in our case) to maximize interpretability and avoid model overfitting. The encoder incorporates Principal Component Analysis (PCA) to compress three-dimensional WRF longwave radiation and shortwave radiation outputs into multiple low-dimensional PCs, each representing the time evolution of a particular radiative anomaly pattern. The PCA reduces the complexity of the data from three-dimensional volumes to 1D time series. The encoder distills knowledge by combining the PCs into traceable compressed representations based on skillful TC intensity forecasts. The scalar output of the encoder is the projection of the radiation structures of individual samples onto the learned time-invariant structures, which roughly represents the spatial similarity between the two structures. We use these scalars to predict the 24-hour, forward-facing ($V_{t+24}-V_{t}$) intensification of TC surface intensity with a linear decoder. A lead time of 24 hours is commonly used for short-term intensity predictions \cite{law2007statistical}. All models presented herein are implemented and trained with the PyTorch deep learning library \cite{paszke2019pytorch}. As shown in Section ~\ref{math:0}, the linear VED framework can be considered a combination of different multiple linear regression (MLR) equations. Training an ML model involves using an iterative procedure called ``stochastic gradient descent`` (SGD) to progressively optimize the weights and biases of the different MLR equations. The SGD procedure for all models is performed with the Adam Optimizer. Whether or not an ML model can reach an optimal solution depends on the ``learning rate'', the step at which the SGD procedure proceeds, and a well-designed loss function. Loss functions are quantifiable measurements that are tied to the optimization process. An example of this is optimizing a model by minimizing a loss function like the mean squared error (MSE) between model predictions and observations. The VED architecture requires a specialized loss function for optimization. The VED architecture can be considered a generalized form of the Variational Autoencoder (VAE) architecture that uses different input-output pairs. Thus, the loss function used to optimize the VED model has the same structure as that commonly used for VAEs:

\begin{equation} 
\label{crossval:VEDloss}
\mathrm{VED loss} = \nu \mathrm{Reconstruction\ Loss} + (1-\nu) \mathrm{KL loss},
\end{equation}

where $\nu\in \left[0,1\right]$ is the weight of the reconstruction loss in the overall VED loss. The reconstruction loss measures how well the model predicts the output (a simple mean absolute error in our case), whereas the KL loss constrains the distribution of the latent space. We use an annealing strategy to train the VED models. The models are first optimized with only the reconstruction loss. We will then continue training the best model with the smallest reconstruction loss with different $\nu$ to optimize the prediction spread. The ``learning rate`` and $\nu$ are model hyperparameters - parameters that define the architecture of an ML model and its training. The best sets of hyperparameters for different data splits are found with the Optuna hyperparameter tuning framework \cite{akiba2019optuna}.

Various aspects of the proposed framework will be elaborated upon in this section. First, we demonstrate that the extracted patterns and their contributions to the VED predictions can be mathematically defined, ensuring full interpretability (Section~\ref{math:0}). Second, we describe methods to incorporate uncertainty quantification in the framework (Section~\ref{methodology:UQ}), which is critical in physical discovery (e.g., for Section \ref{main:result_timeseries}). Finally, we present the metrics used in this work to evaluate prediction skills (Section~\ref{sub:Evaluation}).

\subsection{Mathematical Construction of the Interpretable VED Models} \label{math:0}
As a whole, the main objective of the proposed VED model can be described by a simple equation:

\begin{equation} \label{eq_main:intensification}
\underbrace{\left(\frac{dV_{\mathrm{surf}}}{dt}\right)_{24hr}}_{\mathrm{Intensification}}=b+\underbrace{a_{\mathrm{LW}}\cdot X_{\mathrm{LW}}}_{\mathrm{Longwave\ contribution}}+\underbrace{a_{\mathrm{SW}}\cdot X_{\mathrm{SW}}}_{\mathrm{Shortwave\ contribution}}
\end{equation}
where the prediction target is the temporal change in maximum tangential mean surface winds ($V_{\mathrm{surf}}$) in units of m/s. Radiative contributions to the prediction are proportional to linear regression coefficients $a_{\mathrm{LW}}$ and $a_{\mathrm{SW}}$ (in units of $ \mathrm{m.s^{-1}.K^{-1}}$). The intercept term ($b$; units in $\mathrm{m/s^2}$) in the equation can be thought of as the intensification's expected value over the training set (positive in our case). Intensification rates predictions can be calculated from the sum of radiative contributions ($X_{\mathrm{LW}}$, $X_{\mathrm{SW}}$) that are sampled from normal distributions $\cal N$, with learned means ($\mu_{\mathrm{LW}}$, $\mu_{\mathrm{SW}}$) and learned logarithmic standard deviations ($\mathrm{log} \sigma_{\mathrm{LW}}^2$, $\mathrm{log} \sigma_{\mathrm{SW}}^2$):
\begin{equation}
    X_{\mathrm{LW}}\sim{\cal N}\left(\mu_{\mathrm{LW}},\mathrm{log} \sigma_{\mathrm{LW}}^2\right)\ , \ X_{\mathrm{SW}}\sim{\cal N}\left(\mu_{\mathrm{SW}},\mathrm{log} \sigma_{\mathrm{SW}}^2\right)
\end{equation}
These four statistical moments are constructed with the learned weights and biases in the VED encoder module, and we will show that the conditional means $\mu_{\mathrm{LW}}$ and $\mu_{\mathrm{SW}}$ can be interpreted as projections of the longwave and shortwave fields onto data-driven, three-dimensional patterns. We will further show that the linear regression coefficient and intercept term can be defined with the learned weights and biases for both the encoder and decoder modules. 

\subsubsection{Encoder}
As mentioned previously, Principal Component Analysis (PCA) extracts information from the WRF raw fields. The PCA linearly transforms input physical fields ($X_{i}$) into combinations of orthogonal singular (PCA) spatial modes ($\Pi_{X_{i}}(z,r,\theta)$), and their corresponding time evolution (PC loadings time series; $PC_{X_{i}}(t)$):
\begin{equation} \label{define:PC}
    X_{i}\left(t,z,r,\theta\right)-\overline{X_{i}\left(t,z,r,\theta\right)} = \sum_{i=1}^{N} PC_{X_{i}}(t) \Pi_{X_{i}}(z,r,\theta),
\end{equation}
where $N$ is the number of retained PCA modes. Instead of the raw field, different loadings $PC_{X_{i}}(t)$, which represents the time evolution of different radiative spatial modes, are used as inputs to the VED. These $PC_{X_{i}}(t)$ time series are standardized to have a mean of 0 and a variance of 1 to avoid high variance in the regression weights:
\begin{equation} \label{normalization:PC}
    \widetilde{PC}_{X_i} (t) = \frac{PC_{X_{i}}(t)-\overline{PC_{X_{i}}(t)}}{\sigma_{PC_{X_{i}}(t)}},
\end{equation}
where $\overline{PC_{X_{i}}}$ and $\sigma_{PC_{X_{i}}(t)}$ are the mean and standard deviation of the PC loadings, calculated over the training set.

The second task of the encoder is to combine different PCA modes for longwave radiation and shortwave radiation into two scalars representing the projection of radiation structures onto time-invariant ``mean structures'' ($\Pi_{LW,\mu},\Pi_{SW,\mu}$) and two scalars representing the projection onto ``logarithmic variance structures'' ($\Pi_{LW, log\sigma^2},\Pi_{SW,log\sigma^2} $). These four projection scalars (encoder output) are the linear combinations of the standardized PC loadings (Eq. ~\ref{normalization:PC}),
\begin{equation} \label{MLeq:layer1}
    \begin{aligned}\mu_{LW} & =b_{1,LW,\mu}+\sum_{i=1}^{n_{LW}}a_{1,LW,\mu,i}\times\widetilde{PC}_{LW,i},\\
\log\sigma_{LW}^{2} & =b_{1,LW,\log\sigma^{2}}+\sum_{i=1}^{n_{LW}}a_{1,LW,\log\sigma^{2},i}\times\widetilde{PC}_{LW,i},\\
\mu_{SW} & =b_{1,SW,\mu}+\sum_{i=1}^{n_{SW}}a_{1,SW,\mu,i}\times\widetilde{PC}_{SW,i},\\
\log\sigma_{SW}^{2} & =b_{1,SW,\log\sigma^{2}}+\sum_{i=1}^{n_{SW}}a_{1,SW,\log\sigma^{2},i}\times\widetilde{PC}_{SW,i}.
\end{aligned}
\end{equation}
In these equations, terms with $a$ and $b$ represent the learned weights and biases in the VED framework, respectively. The subscripts in these terms have different options highlighting different aspects of the framework: ``1'' and ``2'' represent learned weights and biases in the encoder module and the decoder module;  ``LW'' and ``SW'' represent the physical variables these coefficients are associated with (longwave and shortwave radiation). Finally, $\mu$ and $\log \sigma^2$ indicate the specific branch these coefficients are in within the VED framework (mean structure and uncertainty structure). 

From Equation \ref{eq_main:intensification}, the scalars defined in Equations ~\ref{MLeq:layer1} will be used to construct normal distributions ${\cal N}\left(\mu,\mathrm{log} \sigma^2\right)$ for longwave and shortwave radiation. Each time the VED model is run, two scalars will be randomly sampled from the two normal distributions and used as inputs in the prediction equation (Eq. ~\ref{eq_main:intensification}). We could quantify the uncertainty of the prediction of TC intensification rates by running the model multiple times with the same inputs. 

\subsubsection{Interpreting the Encoding as a Scaled Projection} \label{math:2}
In addition to guiding how to combine the different PC modes, the learned encoder weights can also be used to obtain the 3-D spatial patterns corresponding to the four scalars (Eq. ~\ref{MLeq:layer1}),
\begin{equation} \label{MLeq:structures}
    \begin{aligned}\Pi_{\mu \mathrm{LW}} & =\lambda_{LW,\mu} \sum_{i=1}^{n_{LW}}a_{1,LW,\mu,i}\times\Pi_{LW_i},\\
    \Pi_{\mu \mathrm{SW}} & =\lambda_{SW,\mu} \sum_{i=1}^{n_{SW}}a_{1,SW,
    \mu,i}\times\Pi_{SW_i},\\
    \Pi_{\mathrm{log} \sigma^2 \mathrm{LW}} & =\lambda_{LW,\mathrm{log} \sigma^2} \sum_{i=1}^{n_{LW}}a_{1,LW,\mathrm{log} \sigma^2,i}\times\Pi_{LW_i},\\
    \Pi_{\mathrm{log} \sigma^2 \mathrm{SW}} & =\lambda_{SW,\mathrm{log} \sigma^2} \sum_{i=1}^{n_{SW}}a_{1,SW,\mathrm{log} \sigma^2,i}\times\Pi_{SW_i},\\.
\end{aligned}
\end{equation}
where $\lambda$ represents scaling factors that ensure the extracted patterns have a norm of 1. The full mathematical deviation of this constant is shown in Appendix A.

But how are the scalars (Eqs. ~\ref{MLeq:layer1}) and the spatial patterns (Eqs. ~\ref{MLeq:structures}) related? In this section, we show that the scalars are scaled projections of raw fields onto the time-invariant patterns. We first introduce an inner product that will help us reinterpret the encoder module: \begin{equation} \label{define:innerproduct}
\left\langle PC_{X_1}|PC_{X_2}\right\rangle_{X_3} = \sum_{i=1}^{N} PC_{X_1}^i PC_{X_2}^i,
\end{equation}
where the $X_3$ subscript indicates that the inner product and the projection are defined with respect to the $X_3$ variable for which the PC decomposition and orthogonal modes are calculated. Using this notation, the longwave conditional mean scalar ($\mu_{\mathrm{LW}}$) can be expressed as the projection of the spatiotemporal longwave heating field onto $\Pi_{\mu\mathrm{LW}}$:

\begin{equation} \label{eq_main:mu_LW}
\underbrace{\mu_{\mathrm{LW}}\left(t\right)}_{\mathrm{Conditional\ Mean}}=\left\langle \underbrace{\mathrm{LW}^{\prime}\left(x,y,z,t\right)}_{\mathrm{Longwave\ Heating\ Anomaly}}\ |\ \underbrace{\Pi_{\mathrm{LW}, \mu}\left(x,y,z\right)}_{\mathrm{Data-Driven\ Pattern}}\right\rangle_{LW}.
\end{equation}

Indeed, the PC loadings $\left(PC_{i,\Pi_{\mu LW}},PC_{i,\Pi_{\mu SW}},PC_{i,\Pi_{\log\sigma^{2}LW}},PC_{i,\Pi_{\log\sigma^{2}SW}}\right) $ of the learned mean ($\Pi_{\mu LW}$, $\Pi_{\mu SW}$) and standard deviation  ($\Pi_{\log\sigma^{2}LW}$, $\Pi_{\log\sigma^{2}SW}$) obey the following equations:
\begin{equation}
\begin{cases} \label{define:innerPCs}
\mu_{LW} = \left\langle LW^\prime\ |\ \Pi_{\mu LW}\right\rangle_{LW}  & =\sum_{i=1}^{n_{LW}}PC_{i,LW}\times PC_{i,\Pi_{\mu LW}},\\
\mathrm{log} \sigma_{LW}^2 = \left\langle LW^\prime\ |\ \Pi_{\log\sigma^{2}LW}\right\rangle_{LW}  & =\sum_{i=1}^{n_{LW}}PC_{i,LW}\times PC_{i,\Pi_{\log\sigma^{2}LW}},\\
\mu_{SW} = \left\langle SW^\prime\ |\ \Pi_{\mu SW}\right\rangle_{SW}  & =\sum_{i=1}^{n_{LW}}PC_{i,SW}\times PC_{i,\Pi_{\mu SW}},\\
\mathrm{log} \sigma_{SW}^2 = \left\langle SW^\prime\ |\ \Pi_{\log\sigma^{2}SW}\right\rangle_{SW}  & =\sum_{i=1}^{n_{LW}}PC_{i,LW}\times PC_{i,\Pi_{\log\sigma^{2}SW}},
\end{cases}
\end{equation}
allowing us to interpret the conditional moments $\left( \mu_{LW},\mu_{SW},\log \sigma^2_{LW}, \log \sigma^2_{SW} \right) $ as projections of raw radiative heating fields onto stationary patterns. Using this projection notation, Equation \ref{eq_main:intensification} may be rewritten as:
\begin{equation} \label{fulleq:goal}
\begin{aligned}
\underbrace{\left(\frac{dV_{surf}}{dt}\right)_{24\mathrm{hr}}}_{\mathrm{Intensification}} & = \underbrace{a_{LW}{\cal N}\left(\left\langle LW|\Pi_{\mu LW}\right\rangle_{LW} \ ,\ c_{LW}e^{d_{LW}\left\langle LW\ |\ \Pi_{\log\sigma_{LW}^{2}}\right\rangle_{LW} }\right)}_{\mathrm{Longwave\ contribution}}\\
 & + \underbrace{a_{SW}{\cal N}\left(\left\langle SW|\Pi_{\mu SW}\right\rangle_{SW} \ ,\ c_{SW}e^{d_{SW}\left\langle LW\ |\ \Pi_{\log\sigma_{SW}^{2}}\right\rangle_{SW} }\right)}_{\mathrm{Shortwave\ contribution}}\\
 & +b,
\end{aligned}
\end{equation}
where the $c_{LW}$, $c_{SW}$, $d_{LW}$, and $d_{SW}$ are constants involved when expanding the variance terms. The definitions of these terms are provided in the next section.

\subsubsection{Effective Weights and Biases in the Prediction Equation} \label{math:3}
In this section, we will define several terms that are left undefined in previous sections. These undefined terms include the ``effective weights'' and bias terms in Equation \ref{eq_main:intensification}, the scaling factor in Equation ~\ref{MLeq:layer1}, and the four variance constants terms in Equation ~\ref{fulleq:goal}. The exact deviation steps for these terms can be found in Appendix A.

The ``effective weights'' terms can be obtained by factoring out the constant terms in the normal distributions after expanding Equation \ref{eq_main:intensification} with Eq. \ref{normalization:PC} and \ref{MLeq:layer1}.

The effective weight of the longwave contribution to TC intensification is,
\begin{equation} \label{effectweight:LW}
a_{LW} = \left|a_{2,LW}\right|\sqrt{\sum_{i=1}^{n_{LW}}\frac{a_{1,LW,\mu,i}^{2}}{\sigma\left(PC_{i,LW}\right)^{2}}},
\end{equation}
whereas the effective weight of the shortwave contribution is
\begin{equation} \label{effectweight:SW}
a_{SW} = \left|a_{2,SW}\right|\sqrt{\sum_{i=1}^{n_{SW}}\frac{a_{1,SW,\mu,i}^{2}}{\sigma\left(PC_{i,SW}\right)^{2}}}.
\end{equation}

The overall model bias ($b$), which can be thought of as the intensification's expected value over the training set (positive in our case), can be obtained by substituting Equation \ref{eq_main:intensification} with Equations \ref{normalization:PC} and \ref{MLeq:layer1} and isolating constant terms that are not in the normal distributions:
\begin{equation} \label{bias}
\begin{aligned}
b =&b_2 + a_{2,LW}\left(b_{1,LW,\mu} - \sum_{i=1}^{n_{LW}} \frac{a_{1,LW,\mu,i} \overline{PC_{i,LW}}}{\sigma(PC_{i,LW})}\right) \\
&+a_{2,SW}\left(b_{1,SW,\mu} - \sum_{i=1}^{n_{SW}} \frac{a_{1,SW,\mu,i} \overline{PC_{i,SW}}}{\sigma(PC_{i,SW})}\right).
\end{aligned}
\end{equation}

This term will be referred to as ``persistence baseline'' in the subsequent sections. Physically, this term can be considered as the main contribution of non-radiative processes to the intensification of all training TCs. 

The scaling factors for the four extracted patterns are mathematically derived by (i) combining equations ~\ref{eq_main:intensification}, ~\ref{MLeq:layer1}, and ~\ref{normalization:PC}, (ii) building upon the equivalence of Equation ~\ref{fulleq:goal} and the output equation from (i), and (iii) assuming the norms (squared) of the structures to be 1. For the mean longwave structure, the scaling factor is,
\begin{equation} \label{equivalence:MEAN_coeff}
\lambda=\left[\left|a_{2,LW}\right|\sqrt{\sum_{i=1}^{n_{LW}}\frac{a_{1,LW,\mu,i}^{2}}{\sigma\left(PC_{i,LW}\right)^{2}}}\right]^{-1}.
\end{equation}
$\lambda$ for the other scalars can be similarly defined by substituting the encoder weights and PC loading time series with those corresponding to the shortwave radiation and the logarithmic variance structures.

Finally, algebraic manipulations allow us to mathematically define the four constants involved in the variance calculation ($c_{LW}, c_{SW}, d_{LW}, d_{SW}$):
\begin{equation} \label{constant:c_lw}
c_{LW}=\sqrt{\frac{|a_{2,LW}|}{a_{LW}}}\exp\left(b_{1,LW,\log\sigma^{2}}-\sum_{i=1}^{n_{LW}}a_{1,LW,\log\sigma^{2},i}\frac{\overline{PC_{i,LW}}}{\sigma\left(PC_{i,LW}\right)}\right),
\end{equation}

\begin{equation} \label{constant:c_sw}
c_{SW}=\sqrt{\frac{|a_{2,SW}|}{a_{SW}}}\exp\left(b_{1,SW,\log\sigma^{2}}-\sum_{i=1}^{n_{SW}}a_{1,SW,\log\sigma^{2},i}\frac{\overline{PC_{i,SW}}}{\sigma\left(PC_{i,SW}\right)}\right),
\end{equation}

\begin{equation} \label{constant:d_lw}
d_{LW}=\sqrt{\sum_{i=1}^{n_{LW}}\frac{a_{1,LW,\log\sigma^{2},i}^{2}}{\sigma\left(PC_{i,LW}\right)^{2}}},
\end{equation}

\begin{equation} \label{constant:d_sw}
d_{SW}=\sqrt{\sum_{i=1}^{n_{SW}}\frac{a_{1,SW,\log\sigma^{2},i}^{2}}{\sigma\left(PC_{i,SW}\right)^{2}}}.
\end{equation}


In the prediction equation, positive longwave contributions to surface wind intensification arise when the values sampled from ${\cal N}\left(\mu_{\mathrm{LW}},\sigma_{\mathrm{LW}}\right) $ are greater than zero. Positive longwave contributions will lead to intensification quicker than the training set reference ($b$), and vice versa. A quicker intensification occurs when the spatial distribution of longwave anomaly relative to the training average projects strongly onto $\Pi_{\mu\mathrm{LW}}$. In contrast, smaller intensification rates correspond to cases in which longwave anomalies and $\Pi_{\mathrm{LW},\mu} $ are orthogonal to each other. Similar logic can be applied to the shortwave contributions.

\subsection{Uncertainty Quantifications for Physical Insights} \label{methodology:UQ}
By setting up a normal distribution based on the learned $\mu$ and $\log\sigma^2$ \cite{kingma2013auto}, the proposed VED enables uncertainty quantification (UQ) for both the extracted structures and the intensification predictions. This section touches upon how we may use VED uncertainty for physical insights.

The ML models are more trustworthy when they provide the full distribution of possible extracted structures and prediction outcomes. The full distributions yield uncertainty information that can be reliably interpreted, which is crucial when we use data-driven techniques to discover new physical processes. For example, we can use prediction uncertainties to assess the relevance of radiation with time. In contrast, uncertainties in the latent structures highlight specific areas in the structure to focus on in future work for scientific discovery. 

One of the potential limitations in the current VED setup, especially when applying to the tropical cyclogenesis problem, is that we predict intensification with only the longwave and shortwave radiation information. It is likely that all pathways to tropical cyclone genesis and intensification are not included in this underdetermined system. We adopted this strategy because we are in a low sample regime, which necessitates restricting the input to avoid overfitting. However, the restricted nature of the model creatively yields physical understanding. Specifically, we argue that the temporal evolutions of model spread and error reveal the changes in the relevance of radiative feedback with time. Large spread and errors should arise when the system is undetermined and requires information from non-radiatively-coupled variables for reliable intensification predictions. These are periods where non-radiative processes are more important predictors of TC intensification. In contrast, we expect smaller model errors and spread when radiative heating is strongly coupled to intensification. 

\subsection{Evaluating the Trained Probabilistic Models\label{sub:Evaluation}}
The trained VED models are evaluated with two criteria -- good mean prediction skills and a well-calibrated uncertainty in the model outputs. We sample the model spread by running each model 30 times and aggregating the 30 model predictions. A suite of stochastic and determinative performance metrics are used to assess the quality of the model. Two stochastic metrics are evaluated: the Continuous Ranked Probability Score (CRPS score) and the Spread-skill reliability (SSREL) value ~\cite{Haynes_etal2023}. 

The CRPS score compares the Cumulative Distribution Function (CDF) of the probabilistic forecasts against the observations; it is also a generalization of the deterministic mean absolute error (MAE) metric for probabilistic models:
\begin{equation} \label{crossval:CRPS}
\mathrm{CRPS}\left(F, y_{true}\right) = \int_{-\infty}^{\infty} \left[F\left(y_{pred}\right)-\mathcal{H}(y_{pred}-y_{true})\right]^2 dy_{pred},
\end{equation}
where $F$ represents the CDF of the model predictions, $\mathcal{H}$ is the Heaviside step function applied to the difference between the truth ($y_{true}$) and one prediction ($y_{pred}$) sampled from the full distribution. A well-calibrated model should have as small a CRPS score as possible.

The SSREL value \cite{Haynes_etal2023} measures the quality of a binned spread-skill plot -- an assessment of the statistical consistency of a probabilistic model \cite{DelleMonoche_etal2013}. A statistically consistent model, sampled from the same distribution as the truth, should have its spread closely match its error. If the spread-skill curve of a model deviates from the 1-1 line, the model is either under-dispersive (overconfident) or over-dispersive (underconfident). The SSREL value measures weighted distances between the model curve and the 1-1 line:
\begin{equation} \label{crossval:SSREL}
\mathrm{SSREL} = \sum_{k=1}^{K} \frac{N_k}{N} \left[\mathrm{RMSE}_k-\overline{\mathrm{SD}_{k}}\right],
\end{equation}
where $K$ is the number of bins, $N_{k}$ is the number of samples in a bin, $N$ is the total number of samples, $\mathrm{RMSE}_{k}$ is the root-mean-square-error of the model predictions for samples within the bin, $\mathrm{SD}$ is the standard deviation of the model predictions. A perfectly calibrated model will have an SSREL value of 0. We report two metrics for the mean deterministic skills: the mean absolute error (MAE) and root mean squared error (RMSE).

\section{Results}

An advantage of our proposed model architecture is that it simultaneously extracts structures relevant to 24-hour intensification rates and the uncertainties in the predictions. Here, we explore using this information to understand (i) how we can use ML to identify temporal periods where radiation is an important driver of intensification, (ii) the relevance of axisymmetric radiation to intensification, (iii) whether asymmetric radiation exerts influence on intensification, and (iv) whether we can show this influence with simple perturbation experiments.

\subsection{Choosing the Best VED Model and Comparison with the Best Baseline Model} \label{subsection: choosing}
\begin{figure*}
    \includegraphics[width=\textwidth]{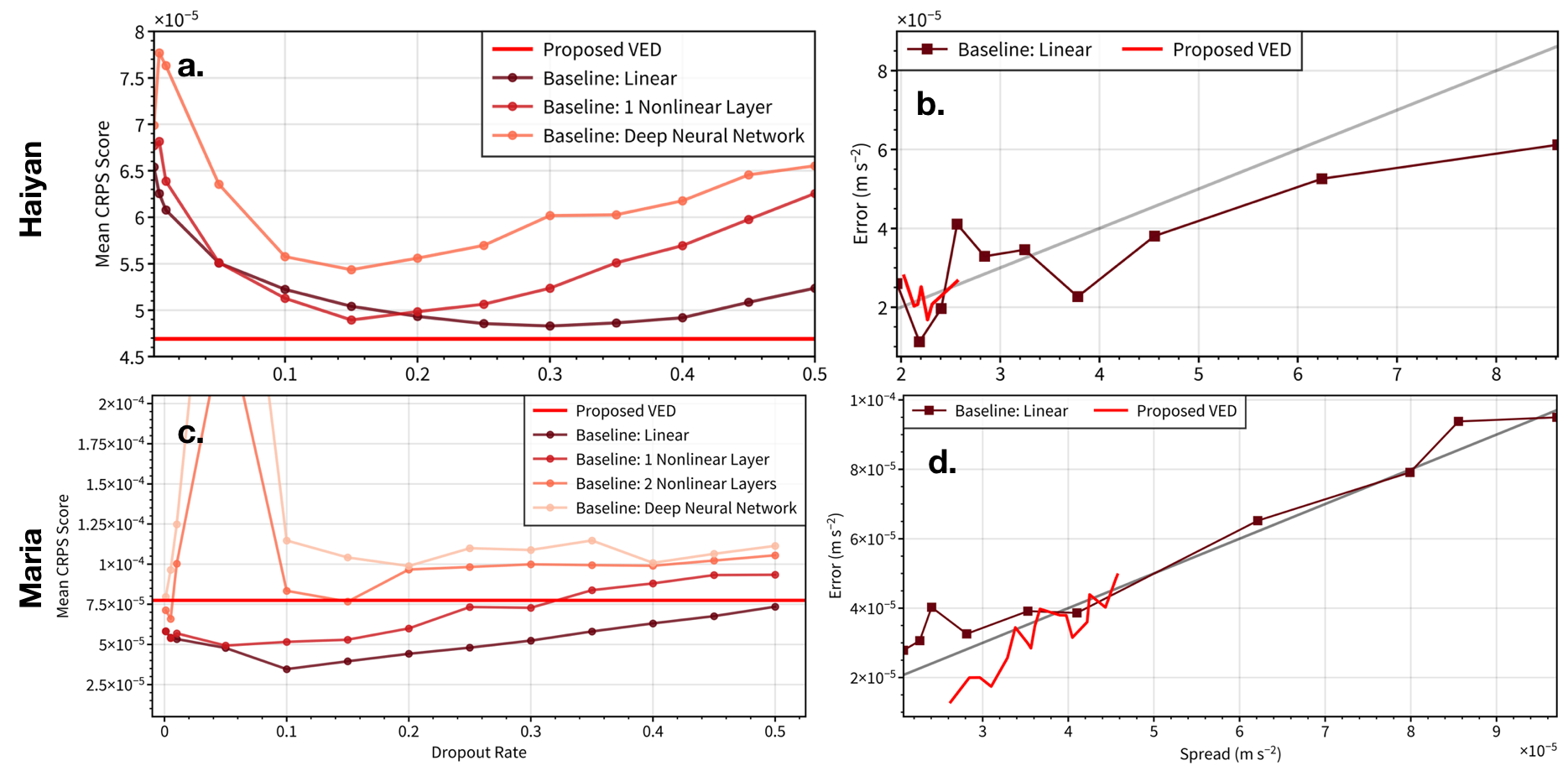}
    \caption{The mean CRPS scores (left column) and the spread-skill diagrams (right column) show that the trained VED models outperform the trained baseline models with different dropout rates and degrees of nonlinearity in the prediction equation (``nonln:3`` represents baseline model with three nonlinear layers in the decoder). Comparing the best fully linear baseline model for Haiyan and Maria (brown lines in panel b and d; $\mathbf{dropout\ rates}$ of 0.3 and 0.1 respectively) and the mean performance of the VED model with the best SSREL score shows that the VED model makes fewer mistakes in its predictions and is generally more well-calibrated than the best baseline model. A well-calibrated model means that most points on the model's spread-skill curve are as close to the 1-1 line (gray lines in b, d) as possible.}  
    \label{fig:validation}
\end{figure*}

It is important to present evidence that the proposed VED model achieves better probabilistic skills than the traditional baseline. Here, we compare the best VED model to a simple Principal Component Regression baseline (description in \ref{SI:PCRegression}). The best VED model is chosen objectively based on the CRPS and SSREL scores. Figure \ref{fig:validation} shows the minimum and median CRPS scores for all trained Haiyan VED and baseline models with different hyperparameters on the validation set. We also substitute the fully linear prediction layer in the baseline model with feed-forward neural networks with different depths to evaluate the responses of prediction skills to nonlinearity.

Adding nonlinearity degrades the median CRPS scores for most baseline models (Fig. \ref{fig:validation}a, c), which justifies keeping the model fully linear. The shape of the CRPS score curves for the baseline models (Fig. \ref{fig:validation}a, c) suggests the existence of an optimal range of $\mathbf{dropout\ rates}$ for better generalizability. For the VED models, a $\lambda$ that is too small, i.e., too large a KL loss during training deteriorates prediction skills.

The CRPS score comparison above establishes optimal ranges for the $\mathbf{dropout\ rates}$ (the baseline models) and $\lambda$ (the VED models). The best models for comparison are determined by calculating the SSREL scores for all models trained with these optimal coefficients. The spread-skill plot for the best baseline and VED models (Fig. \ref{fig:validation}) provides a strong justification for using the VED model in our study. Specifically, the spread-skill curve of the VED model is much shorter than the baseline one, which indicates the VED predictions are more accurate. Compared to the best baseline model, the best VED model better captures the peaks in the training dataset and removes the large biases in early intensification rates seen in the baseline predictions on the test set. Based on these comparisons, we conclude that the VED model is superior to the baseline model for our research task.

For Maria, the best VED model performs similarly to the best baseline model in terms of the minimum CRPS score (Fig. \ref{fig:validation}c). The uncertainty for both the best baseline model and best VED model are well calibrated. However, the VED model is again preferable for Maria because of the smaller prediction errors (Fig.  \ref{fig:validation}d). 

Interestingly, the benefit of the VED model compared to the baseline model seems to scale to the sample size. The VED model always overperforms the baseline for the Haiyan ensemble case with a larger sample size (Table \ref{table:SIhaiyan}), whereas the VED model mostly only overperforms the baseline in probabilistic metrics for the Maria simulations (Table \ref{table:SImaria}). A potential explanation for worse VED skills for Maria is that the more complex model (VED) overfits the training data in cases with low sample size.

\begin{table}
\centering
\caption{Prediction skills of the best VED and the best baseline model on the Haiyan ensemble. Also shown in the table is the median of the prediction skills distribution of all trained models with different hyperparameter settings (numbers in brackets). All values in the table are multiplied by $10^5$ for readability. The best model is indicated with bolded numbers.}
\begin{tabular}{l|l|l|l|l}
Experiment                & Metric & Training    & Validation & Test \\ \hline
\multirow{4}{*}{VED}      & CRPS   & \textbf{2.83} (3.35) &  \textbf{1.65} (4.69)          &  \textbf{2.86} (3.73)    \\
                          & SSREL  & \textbf{1.31} (2.61) &  \textbf{0.33} (3.54) &  \textbf{1.03} (2.44)    \\
                          & RMSE   & \textbf{4.52} (5.34) &  \textbf{2.61} (6.99) &  \textbf{4.83} (5.95)    \\
                          & MAE    &  \textbf{3.67} (4.25) &  \textbf{2.15} (5.76) & \textbf{3.75} (4.74)     \\ \cline{1-1}
\multirow{4}{*}{Baseline} & CRPS   &  3.23 (3.48) &  2.26 (4.83) & 3.45 (3.90)     \\
                          & SSREL  &  1.89 (2.45) &  0.67 (4.19) & 1.17 (2.76) \\
                          & RMSE   &  5.35 (5.68) &  3.67 (7.17) & 5.54 (6.09) \\
                          & MAE    &  4.28 (4.56) &  2.96 (6.00) & 4.31 (4.93)  \\ \hline
\end{tabular}
\label{table:SIhaiyan}
\end{table}

\begin{table}
\centering
\caption{Prediction skills of the best VED and the best baseline model on the Maria experiments.}
\begin{tabular}{l|l|l|l|l}
Experiment                & Metric & Training    & Validation & Test \\ \hline
\multirow{4}{*}{VED}      & CRPS   & 1.76 (5.88) &  \textbf{2.28} (7.78)          &  \textbf{1.39} (5.26) \\
                          & SSREL  & 0.59 (3.51) & \textbf{0.80} (4.33) &  \textbf{0.33} (2.95) \\
                          & RMSE   & 3.74 (9.27) &  3.93 (10.12) &  \textbf{2.21} (7.32) \\
                          & MAE    &  2.21 (7.95) & 3.20 (9.95) & \textbf{1.74} (7.08) \\ \cline{1-1}
\multirow{4}{*}{Baseline} & CRPS   &  \textbf{1.16} (2.32) &  2.39 (3.46) & 2.32 (2.60) \\
                          & SSREL  &  \textbf{0.36} (1.62) &  0.88 (3.28) & 1.68 (2.21) \\
                          & RMSE   &  \textbf{2.46} (4.57) &  \textbf{3.57} (6.22) & 3.54 (3.84) \\
                          & MAE    &  \textbf{2.03} (3.21) &  \textbf{3.13} (4.85) & 3.22 (3.51)  \\ \hline
\end{tabular}
\label{table:SImaria}
\end{table}

\subsection{Prominence of Radiative Feedbacks in the Early Intensification Phase} \label{main:result_timeseries}
\subsubsection{Identifying Periods of Radiatively-Driven Intensification}

By construction, our simple VED model predicts TC intensification exclusively from radiative heating, overlooking significant contributions from surface fluxes and wind-induced surface heat exchange  \cite{ZhangEmanuel2016, MurthyBoos2018}. Recent modeling studies \cite<e.g., >{Smith_etal2020, Yang_Tan2020} suggest that radiative heating could be less critical to TC intensification beyond the initial genesis or spin-up stage.

To investigate whether this holds in our case studies, we use the VED model to \textit{identify periods when radiative feedbacks dominate TC intensification}. Instances with significant model errors or uncertainties are times when radiative heating alone cannot predict TC intensification. In contrast, instances with accurate predictions and minimal uncertainty are likely times when radiative heating is dominant.

\begin{figure*}
    \includegraphics[width=\textwidth]{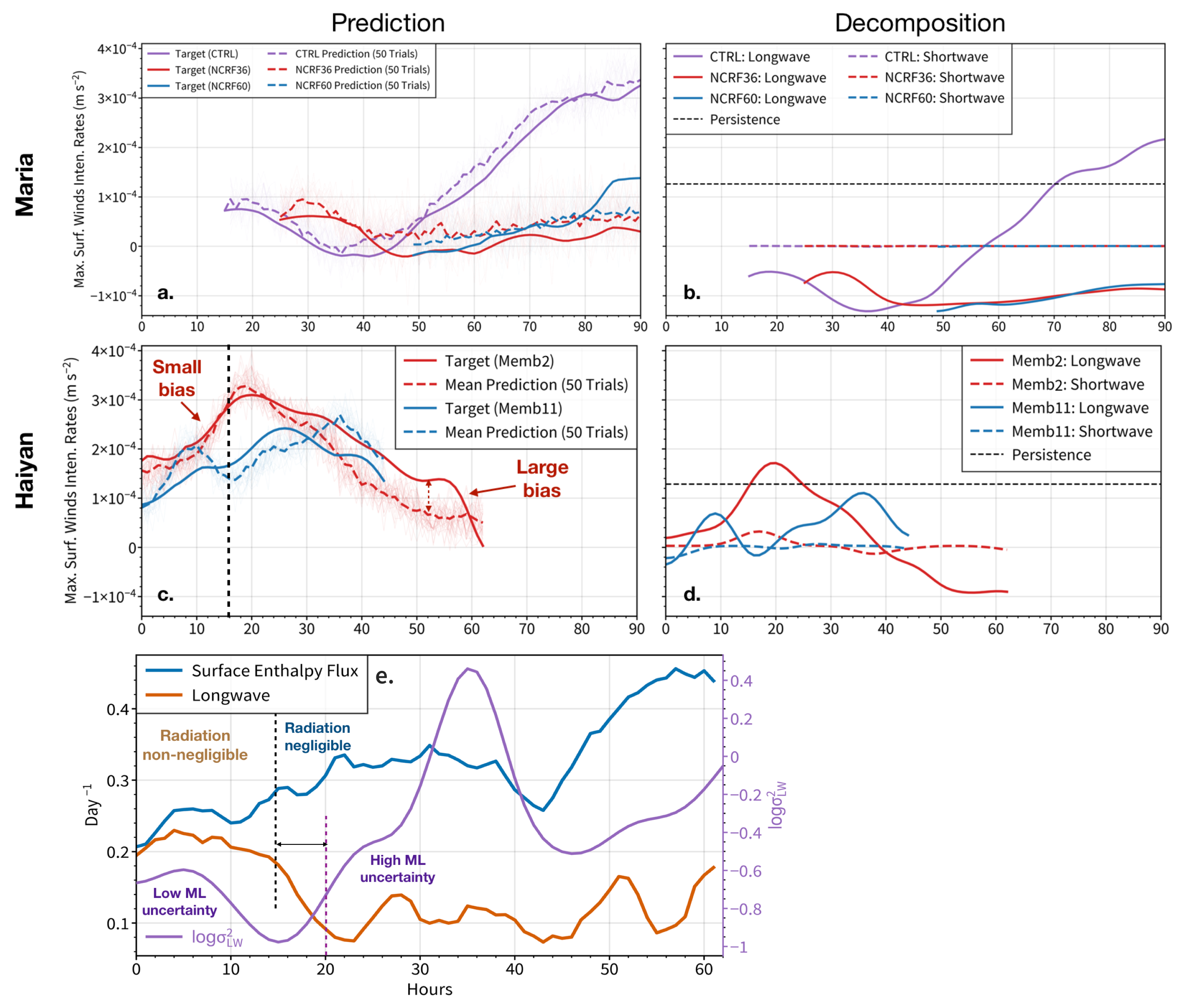}
    \caption{Decomposing tropical cyclone intensification predictions for Maria and Haiyan shows longwave radiative heating's link to early intensity differences. We present mean VED surface intensification predictions (dashed) for Maria (a) from three WRF simulations and for Haiyan (c) from two ensemble members, with actual rates of intensity change (thick). The shadings in the left columns illustrate the range of possible VED predictions given the same inputs using a Monte Carlo approach. Panels (b) and (d) show longwave and shortwave radiation contributions to the mean VED predictions. Zooming in on Haiyan Member 2, Panel (e) compares the evolution of VED uncertainty associated with longwave radiation (purple) against two MSE variance sources (blue and brown). In Maria's mechanism-denial experiment and the early phase of Haiyan, the model associates the difference between a quickly intensifying TC and a slowly intensifying one with longwave radiation. The vertical dashed lines in panel (e) show the two-stage behavior in the longwave MSE variance source term (black) and similar behavior in VED prediction uncertainties (purple). The time lag between the two methods is shown in panel (e; black arrow).\label{fig:prediction_uncertainty}}
\end{figure*}

Our VED's capability to distinguish radiative heating-dominated stages is crucial for reliability assessment and scientific discovery. Figure~\ref{fig:prediction_uncertainty} presents the mean prediction and prediction spread of the best-calibrated VED models for Haiyan and Maria. For Maria, the probabilistic ML models replicate the intensification rate reduction in CRF mechanism-denial experiments (Fig.~\ref{fig:prediction_uncertainty}a). Decomposing predicted intensification into longwave and shortwave contributions (Fig.~\ref{fig:prediction_uncertainty}b) shows that the slower intensification in the mechanism-denial experiments is primarily attributable to the longwave component. This result is reassuring as it identifies the longwave component of CRF as the main contributor to the early intensification of TCs, consistent with the ``cloud greenhouse effect'' framework in \citeA{Ruppert_etal2020}. However, the model underestimated the intensification rate at the latter stage of the NCRF-60h TC's life cycle, possibly due to unaccounted non-radiative processes like the surface fluxes feedback \cite<e.g., >{ZhangEmanuel2016}. Furthermore, the model assigned a smaller weight to the shortwave channel in the linear prediction equation for Maria (Fig.~\ref{fig:prediction_uncertainty}b), resulting in a shortwave contribution close to zero. We believe the model has learned the strong effect of disabling CRF in the sensitivity experiments and its impact on intensification, which manifests mainly in the longwave channel. The shortwave contribution is larger in the realistic Haiyan ensemble simulations (Fig.~\ref{fig:prediction_uncertainty}d).

For Haiyan, where CRF always exists, we compare the VED predictions for a high intensification rate Haiyan ensemble member (Member 2) to those for a slow intensification member (Member 11) to assess the role of radiation in realistic conditions (Fig.~\ref{fig:prediction_uncertainty}c). Increased bias (Fig.~\ref{fig:prediction_uncertainty}c) and wider uncertainty range (Fig.~\ref{fig:prediction_uncertainty}e) for Member 2 predictions in the latter part of the TC's life cycle shows the limitation of the VED model to understand the mature phase of TC intensification. The prediction distribution for samples taken during the high uncertainty phase is close to the training data distribution (prior), implying the need for non-radiative inputs to adequately constrain the probabilistic predictions and reduce overall model bias. In contrast, the model predictions for samples taken from the early intensification phase of Member 2 are accurate, indicating predictability from the radiative heating fields.

Comparing linear decompositions of VED predictions (Fig.~\ref{fig:prediction_uncertainty}d) reveals that lower intensification rates for Member 11 (Hours 10-20) are due to reduced longwave contribution. Positive longwave contribution in Member 2 leads to a faster intensification rate beyond the persistence baseline (overall model bias). In the next section, we analyze the radiation structures in the two ensemble members to provide physical interpretations of how differences in radiative heating structures might explain the differing intensification rates of the two ensemble members.

\subsubsection{Qualitative Agreement with Existing Diagnostic Tools}
In the previous section, we claimed that transitioning from an accurate, low-uncertainty regime to an inaccurate, high-uncertainty regime in the VED predictions can be understood as separating periods where radiation is more critical from non-radiative processes are more important. Here, we use well-established budget analysis tools to evaluate the extent to which the claim is valid.

We can analyze the cyclogenesis process with the Frozen Moist Static Energy Spatial Variance Budget \cite{WingEmanuel2014,muller2018acceleration,Carstens2020TropicalCF}:
\begin{equation} \label{eq:MSE_spatialvariance_budget}
\frac{1}{2}\frac{\partial \mathrm{var} \hat{h}}{\partial t} = \underbrace{\overline{\hat{h}^{\prime} \mathrm{LHF}^\prime} + \overline{\hat{h}^{\prime} \mathrm{SHF}^\prime}}_{\mathrm{SEF\ Contribution}} + \underbrace{\overline{\hat{h}^{\prime} \mathrm{NetLW}^\prime} + \overline{\hat{h}^{\prime} \mathrm{NetSW}^\prime}}_{\mathrm{Radiative\ Contribution}} - \overline{\mathbf{\nabla}_h \cdot \vec{u} \hat{h}},
\end{equation}

\noindent where $\mathrm{var} \hat{h}$ is the spatial variance of vertically integrated moist static energy (MSE), and $\mathrm{SEF}$ (the Surface Enthalpy Flux) is the sum of $\mathrm{LHF}$ (the Latent Heat Flux) and $\mathrm{SHF}$ (the Sensible Heat Flux). The radiative contribution consists of net column longwave radiative flux convergence ($\mathrm{NetLW}$) and net column shortwave radiative flux convergence ($\mathrm{NetSW}$). Primes indicate anomalies relative to the mean of the spatial domain, represented by overlines. MSE spatial variance source terms are obtained by spatially averaging all terms on the right-hand side of Equation ~\ref{eq:MSE_spatialvariance_budget}. We perform the spatial averaging from 0 to 600 km from the TC center. The MSE variance summarizes the spatial distribution of frozen MSE surrounding a developing tropical cyclone, with the biggest contribution coming from moisture \cite{CarstensWing2022}. Since tropical cyclogenesis shares similarities with rotating convective self-aggregation, TCs form as the TC thermodynamics transition to an aggregated state, characterized by a compact moisture blob surrounded by drier air \cite<positive $ \partial_{t}\mathrm{var}\hat{h}$, e.g., >{muller2018acceleration,Carstens2020TropicalCF}. From Equation \ref{eq:MSE_spatialvariance_budget}, the source terms for the MSE variance are the covariance between the existing MSE anomalies and different flux anomalies. Creating positive MSE variance anomalies necessitates spatially-aligned anomalies. For example, positive radiative contribution to MSE variance may arise from warm longwave heating anomalies in the high energy TC inner core or cool radiative anomalies in the drier TC surroundings \cite{Nicholls2015}. Inward moisture transport from radiatively-driven secondary circulations \cite<in-up-out circulations; e.g., >{Nicholls2015, Ruppert_etal2020, Smith_etal2020} can enhance MSE variance by redistributing moisture towards the TC center.

Fig~\ref{fig:prediction_uncertainty}e compares the time evolution of the radiation and surface enthalpy flux MSE variance source terms and the time evolution of the learned logarithmic variance structure for the longwave channel ($\log \sigma^2_{LW}$). The advection term is not considered due to the coarse temporal resolution of the saved WRF outputs. Focusing on Haiyan member 2, the contribution from $\mathrm{NetLW}$ to the overall MSE variance budget is initially comparable to the contribution from $\mathrm{SEF}$ but becomes less important towards the end of the time series. This behavior is consistent with the expected increase in prominence of surface fluxes feedback ~\cite<WISHE; >{Emanuel1986, RotunnoEmanuel1987, ZhangEmanuel2016} after the initial genesis stage. Consistent with the MSE variance budget analysis, we see two distinct phases in the $\log \sigma^2_{LW}$ time series, one with smaller values before ~20h (fewer prediction uncertainties) and one with larger values after ~20h (more prediction uncertainties). It is encouraging that the $\mathrm{NetLW}$ term and $\log \sigma^2_{LW}$ time series have a two-stage behavior as it ensures that the ML model is trustworthy and has learned physically meaningful relationships. 

A notable caveat to the above discussion is that our ML framework predicts TC kinematic changes, whereas the MSE variance budget measures thermodynamical changes. This distinction potentially explains a 5-hour time lag between the decrease in longwave contribution to MSE variance and the increase in ML prediction uncertainties. The idea that thermodynamic forcing precedes kinematic changes can be supported by Figure 2 in \citeA{Ruppert_etal2020}, where TC surface intensification occurs 12-24 hours after the drop in SEF contributions. The analysis in \citeA{tang2017coupledforcingTCG} also shows that strong moist entropy forcing precedes the genesis time of idealized axisymmetric TCs. Finally, airborne observations suggest that the TC core becomes and remains close to saturation for some time before the build-up of storm circulations \cite{bell2019mesoscaleKarl}. 

\subsection{Axisymmetric results: The dominance of upper-level longwave radiation} \label{main:result_axisymmetric}
The proposed framework's physical interpretability relies on the model predictions and the extracted structures. In the following sections, we progressively highlight different aspects of the extracted structures to show how they can clarify the role of radiation in TC intensification. We start by analyzing the azimuthal mean of VED-extracted structures to illustrate how spatial gradients in radiative anomalies affect TCs. We obtain these structures by multiplying the PCA spatial modes and the trained encoder weights.

\subsubsection{Maria}
\begin{figure*}
    \includegraphics[width=\textwidth]{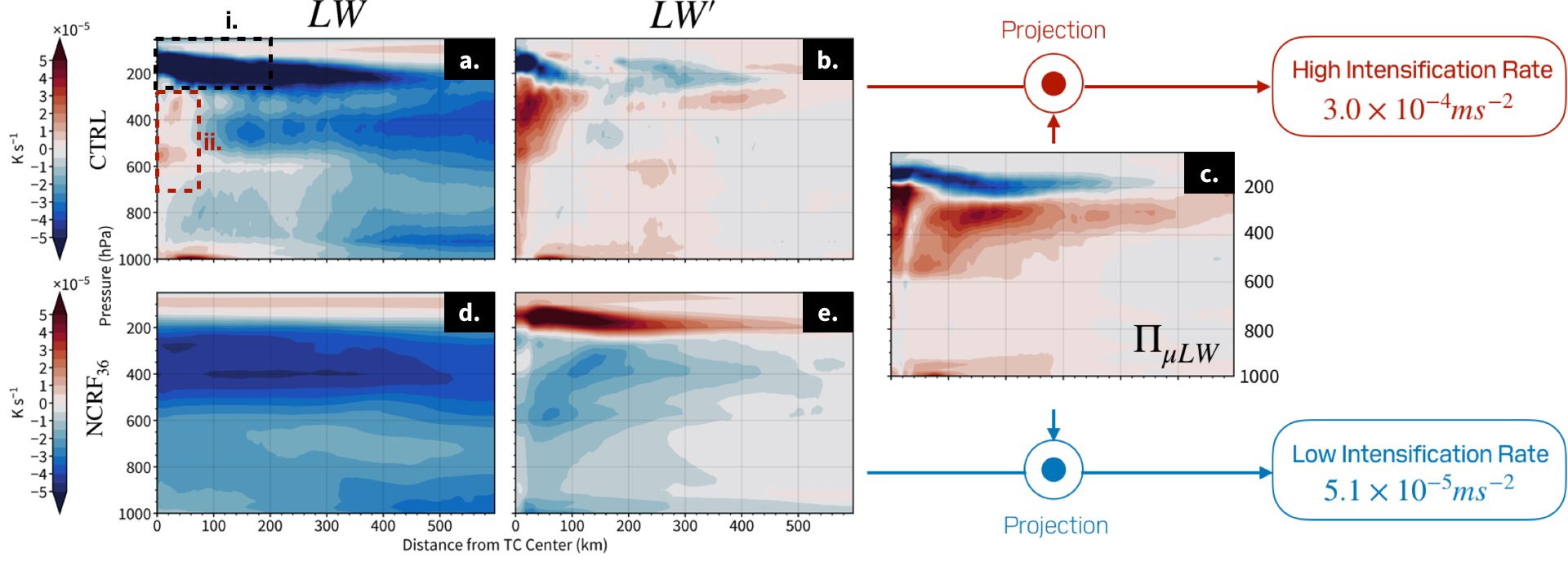}
    \caption{The learned data-driven structure for the mean longwave prediction ($\Pi_{\mu\mathrm{LW}}$; c) for Maria shows a prominent upper-level longwave anomaly dipole. A sample from Hour 80 of the CTRL simulation (a-b), which is predicted by the ML model to have a high intensification rate because of a positive projection of $LW^\prime$ (b; perturbation compared to the training mean) onto $\Pi_{\mu\mathrm{LW}}$ (c). We contrast the CTRL sample with another sample taken at the same hour from the NCRF-36h simulation (d-e), which is predicted to have a negative intensification rate due to a negative projection of $LW'$ (e) onto $\Pi_{\mu\mathrm{LW}}$. The two samples illustrate the physical meaning of the projections: a positive projection (red arrows) occurs when the $LW'$ is similarly spatially to $\Pi_{\mu\mathrm{LW}}$, whereas a negative projection occurs when the $LW'$ is opposite in sign to $\Pi_{\mu\mathrm{LW}}$.}
    \label{fig:latent_structures_maria}
\end{figure*}

In the case of the Maria simulations (Figure ~\ref{fig:latent_structures_maria}), the VED model extracts a $\mu_{LW}$ pattern with an upper-level longwave anomaly dipole and a shallow cloud radiative signal near the surface (Figure ~\ref{fig:latent_structures_maria}c). The anomaly fields are defined with respect to the training mean. In other words, it shows how the radiative heating structure of a sample deviates from the mean of all training samples. This way of defining the anomaly also eliminates the need to recalculate the PCs for individual experiments. To demonstrate how the learned $\Pi_{\mu\mathrm{LW}}$ (Fig. ~\ref{fig:latent_structures_maria}c) encodes physical knowledge, we compare Hour 80 from the Maria CTRL simulation (Fig. ~\ref{fig:latent_structures_maria}a, b) to a sample taken from the same time in the NCRF-36h experiment (Fig. ~\ref{fig:latent_structures_maria}d, e). The VED framework predicts higher surface intensification rates for the CTRL sample due to positive $\mu_{LW}$. In contrast, the NCRF-36h sample has a lower predicted intensification rate due to negative $\mu_{LW}$. Comparing Figure ~\ref{fig:latent_structures_maria}b and c, we see that a positive $\mu_{LW}$ arises when the anomaly field is spatially distributed in the same way as $\Pi_{\mu\mathrm{LW}}$. In the raw longwave radiation field (Fig. ~\ref{fig:latent_structures_maria}a), the positive $\mu_{LW}$ of the CTRL sample corresponds to the concentration of strong longwave cooling near the cloud top (i in Fig. ~\ref{fig:latent_structures_maria}a) and heating near the TC center (ii in Fig. ~\ref{fig:latent_structures_maria}a). In contrast, the negative $\mu_{LW}$ in the mechanism denial example features a lowered and weakened longwave cooling and the absence of heating near the TC center; both contribute to a weaker simulated TC. From the Maria example, we demonstrate that combining the sign of the projection ($\mu_{LW}$) and the raw fields provides valuable information on why NCRF-36h TC fails to intensify. The next section uses $\mu_{LW}$ to understand why two Haiyan ensemble members have different intensification rates during their organization phases.

\subsubsection{Haiyan}
\begin{figure*}
    \includegraphics[width=\textwidth]{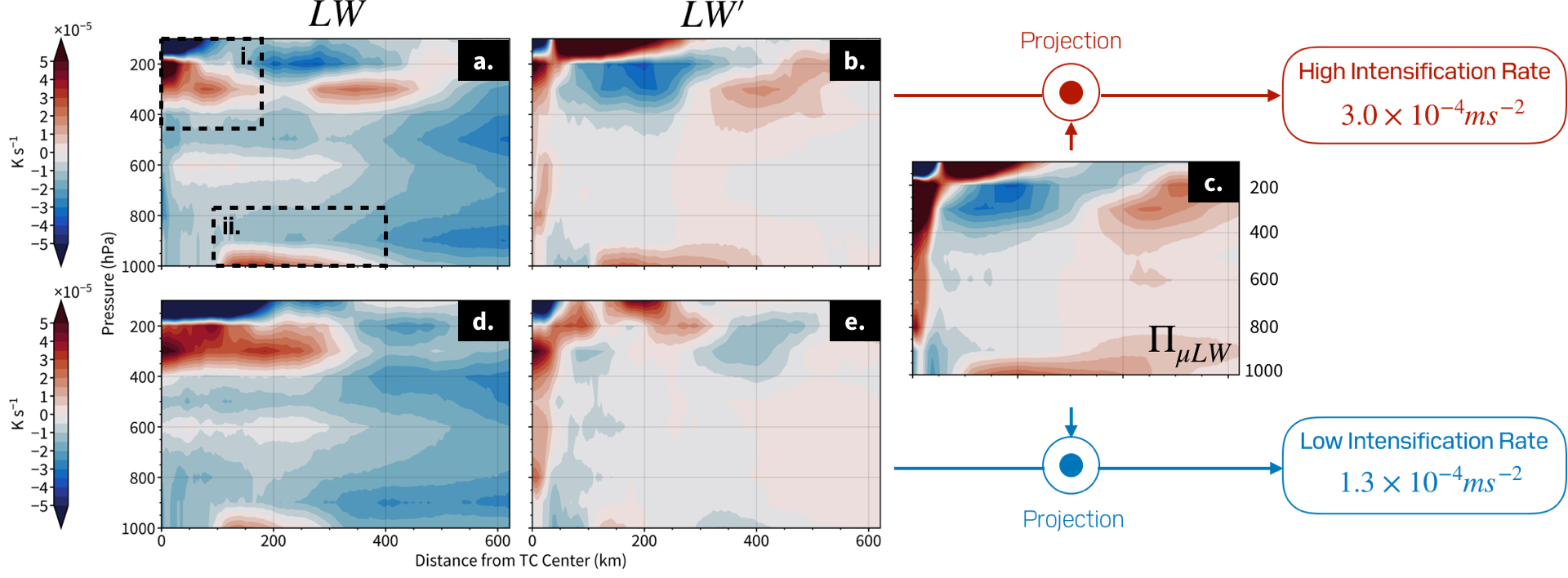}
    \caption{The learned data-driven structure for the mean longwave prediction ($\Pi_{\mu\mathrm{LW}}$; c) for Haiyan shows the spatial distribution of $LW^\prime$ that is most correlated with early TC intensification. The example that is predicted to have a high intensification rate (a,b) is taken from Hour 15 of Member 2, whereas the predicted low intensification rate (d,e) example is taken from Hour 17 of Member 11. The Member 2 $LW^\prime$ example has a strong positive projection (red arrows) onto $\Pi_{\mu\mathrm{LW}}$ (c), which is indicative of concentrated inner-core deep convection (200 hPa longwave anomaly dipole within 100 km of TC center) in the raw azimuthal-averaged $\mathrm{LW}$. Additionally, the Member 2 example also features shallow clouds in the outer core (900 hPa anomaly dipole between 100-300 km from the TC center; b). Both the inner-core deep convection and outer-core shallow cloud signatures are weaker in the Member 11 example, which leads to a lower predicted intensification rate (blue arrow).}
    \label{fig:latent_structures}
\end{figure*}

The azimuthal mean of $\mu_{LW}$ for Haiyan (Fig. ~\ref{fig:latent_structures}c) exhibits a vertical dipole pattern around 200 hPa and a shallow vertical dipole at 900 hPa, reflecting broad anvil clouds in the outer core and shallow clouds in the inner core. We compare samples from two members: Member 2, with a significant positive $\mu_{LW}$ (larger predicted intensification rate), and Member 11, with a small $\mu_{LW} $ (smaller predicted intensification rate). In the Member 2 sample, positive $\mu_{LW}$ indicates weakening of the upper-level longwave dipole between 50 km to 300 km from the TC center and a vertically expanded upper-level longwave heating near the TC center (i in Fig. ~\ref{fig:latent_structures}a), along with a more prominent 900 hPa shallow cloud radiative dipole (ii in Fig. ~\ref{fig:latent_structures}a). These longwave patterns may indicate deep convective development, rising outflow height near the TC center, destabilized inner core upper-level thermal stratification, and enhanced shallow cloud frequency. Colder upper tropospheric temperatures and higher outflow layers have been shown to boost TC intensity. Following balanced dynamics \cite{Eliassen:1952,PendergrassWilloughby2009}, upper-level radiative cooling triggers secondary circulations that accelerate surface tangential winds in idealized TCs \cite{Trabing_etal2019}. From an energetic perspective, rising outflow layers enhance the thermal efficiency of a TC heat engine, resulting in a stronger TC \cite{Ramsay2013,Wang_etal2014}. While shallow clouds are suggested to assist cyclogenesis by moistening the lower troposphere and spinning up the near-surface circulation \cite{Wang2014}, our VED model analysis implies their effect on the overall TC intensification is relatively minor. Decomposing the model prediction by vertical level (description in \ref{appendix_VL}) suggests that the shallow cloud contribution to intensification is between two and one order of magnitude smaller than the upper-level radiative contribution (Fig. \ref{fig:5:sensitivity}b). These findings corroborate those from idealized simulations \cite<e.g.,>{Kilroy2021}.

\subsection{Asymmetric Radiative Heating favors Tropical Cyclone Intensification} \label{main:result_asymmetric}
\begin{figure*}
    \includegraphics[width=\textwidth]{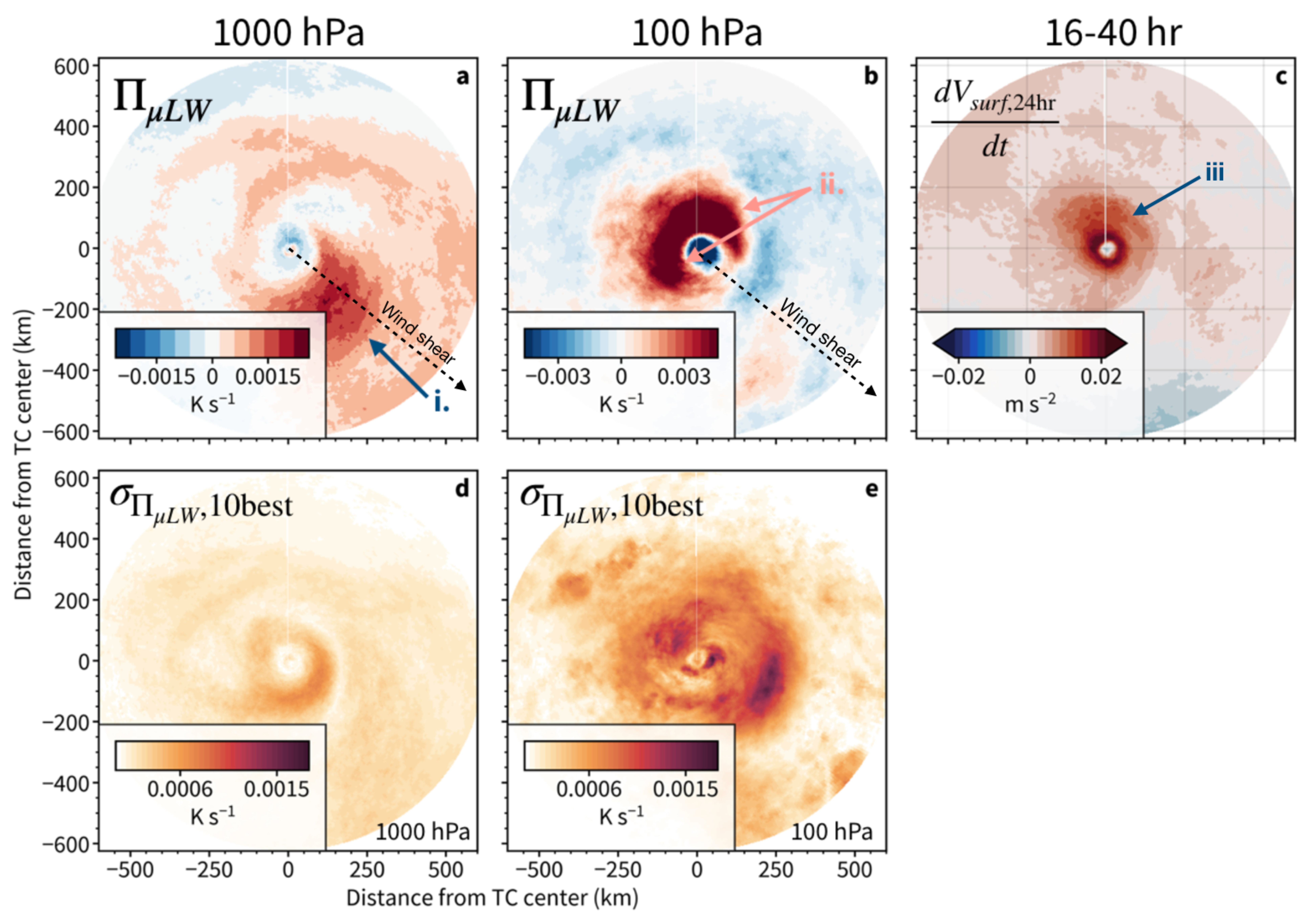}
    \caption{The spatial cross-sections of the best-performing model's mean longwave pattern $\Pi_{\mu\mathrm{LW}}$ at (a) 1000 hPa and (b) 100 hPa both exhibit substantial wavenumber-1 asymmetry in the inner core (0-200 km from the TC center). The trustworthiness of individual anomaly areas in $\Pi_{\mu LW}$ is assessed by the standard deviation of the 10 best models' $\Pi_{\mu\mathrm{LW}}$ at (c) 1000 hPa and (d) 100 hPa. We highlight two longwave anomalies that are trustworthy (small $\Pi_{\log \sigma^2 \mathrm{LW}}$): a low-level shallow cloud signature (i in a) and the deep convection downwind of the shallow cloud signature (ii in b). We compare (c) the 24-hr surface wind intensification associated with the Member 2 example in Figure~\ref{fig:latent_structures}a-b to show a strong spatial correlation between the broad surface wind intensification (iii in c) and the upper-level longwave anomaly signature (ii in b).}
    \label{fig:4}
\end{figure*}

Here, we show that some asymmetric longwave radiative anomaly structures are potential predictors for intensification. The linear combinations of PC longwave eigenvectors yield a complex, spatially asymmetric, three-dimensional radiative heating pattern (Fig. ~\ref{fig:4}a,b). The $\Pi_{\mu\mathrm{LW}} $ cross sections at 1000 hPa (Fig. ~\ref{fig:4}a) and 100 hPa (Fig. ~\ref{fig:4}b) both exhibit distinct wavenumber-1 asymmetry, with a shallow cloud radiative signature at 1000 hPa (i. in Fig. ~\ref{fig:4}a; downshear right quadrant) distributed upwind of the deep convective signature at 100 hPa (ii. in Fig. ~\ref{fig:4}b; downshear left quadrant). Considering how the secondary circulations associated with CRF may intensify tangential winds and spin up surface cyclones \cite{Ruppert_etal2020}, we postulate that any positive longwave contribution to Member 2 surface winds should be distributed mainly in the TC's northern half. We validate this with the surface wind intensification from Hour 16 to 40 (Fig. \ref{fig:4}c), which shows the effect of a positive $\mu_{LW}$. The surface wind acceleration is indeed broader over the TC's northern half, with the strongest acceleration located downwind of the downshear left positive upper-level $\Pi_{\mu\mathrm{LW}}$ with minimal uncertainty. The spatial correlation between surface wind acceleration and $\Pi_{\mu\mathrm{LW}}$ supports the hypothesis that upper-level wavenumber-1 longwave heating anomaly contributes positively to surface intensification. While shallow clouds have a minor contribution to the overall TC intensification (Section 4.2), the shallow cloud radiative signature (i in Fig. ~\ref{fig:4}a) still coincides spatially with the intensifying \textit{local} winds in the downshear right quadrant. 

To identify which mesoscale anomalies in $\Pi_{\mu\mathrm{LW}}$ are most unambiguously correlated to intensification, we create standard deviation maps from the ten best-performing models' $\Pi_{\mu\mathrm{LW}}$ structures (Fig. ~\ref{fig:4}d,e). Low standard deviations suggest that the ML models consistently find the exact relationship between mesoscale anomalies and intensification in a specific area. We should prioritize such areas in our analysis. Using the deviation maps at 100 hPa (Fig. ~\ref{fig:4}e), we conclude that the 100 hPa $\Pi_{\mu\mathrm{LW}}$ positive anomaly in the downshear left quadrant (i in Fig. ~\ref{fig:4}a) is unambiguously related to intensification (low uncertainty). Asymmetric deep convection is typically located in the downshear quadrants of sheared TCs \cite<e.g., >{rios2024review,carstens2024tropical}. In contrast, the upper-level anomalies in the upshear quadrants have more uncertainty. The higher uncertainty in these quadrants may reflect non-shear mechanisms not directly related to intensification on precipitation asymmetry, such as TC movement \cite{tu2022increase}. Finally, the 1000 hPa deviation map (Fig. ~\ref{fig:4}d) shows model consistency in depicting a cold LW anomaly at the TC center and a shallow cloud signature in the downshear right quadrant (i in Fig. ~\ref{fig:4}a). A high standard deviation in the boundary between the two areas indicates uncertainty in the spatial extent of these signatures across models (Fig. ~\ref{fig:4}d).

Considering these results, the longwave anomaly signature associated with deep convection in the downshear left quadrant is the feature that has the strongest correlation to surface intensification. Widespread inner core convective development near and slightly downwind of the upper-level longwave anomalies (not shown) in the Haiyan Member 2 example points to a link between longwave anomalies, secondary circulations, and deep convective development, which also facilitate the axisymmetrization of TC structures - an indicator of TC genesis \cite{hendricks2004role} and intensification \cite{shimada2017tropical}.

\section{Discussion}

\subsection{Complementing Physics-Based Budget Analysis with Machine Learning Diagnostics}
This section briefly discusses how the proposed ML-based method complements existing physics-based diagnostic methods. The first category of physics-based diagnostic methods is budget equations. Budget equations are attractive in that each term in the equation has a well-defined physical meaning; the equations suitable for the tropical cyclogenesis problem include the moist static energy variance budget \cite{WingEmanuel2014}, horizontal momentum budget \cite{huang2018concentric}, kinematic energy budget \cite{wang2016kinetic}, and available potential energy budget \cite{novak2018local}. One disadvantage of budget analyses for our problem is that they do not directly predict surface intensification from thermal forcing. The MSE variance budget calculates the spatial variance in MSE (a thermodynamic term), whereas the horizontal momentum equation lacks a thermal source term. There is no direct equation that links the MSE variance and TC intensity, even though a statistical correlation exists between the two \cite{wing2022acceleration}. While it is possible to get physical insights by comparing the thermodynamic and kinematic budget terms side-by-side, such analyses are qualitative, not quantitative. Another disadvantage common to all budget analysis methods is that they rely on variables not typically included in simulation output lists. It is well-known that these budgets are hard to close \textit{post-hoc} \cite{chen2020towards}, i.e., calculating the budget terms with typical model outputs, especially if the model outputs are stored infrequently. A better option is to estimate those terms directly in the model (online). However, this requires rerunning pre-existing large cloud-resolving simulation datasets with the online calculation of budget terms \cite<e.g., >{stevens2019dyamond}, which is often computationally prohibitive. Hence, our ML framework complements physics-based budgets by (i) establishing quantitative relationships between thermodynamic forcing and kinematic changes and (ii) relaxing the data requirements in temporal frequency and online budget calculations. 

The second category of physics-based diagnostics used in tropical cyclogenesis studies is the Sawyer-Eliassen Equation \cite<SEQ; >{PendergrassWilloughby2009}. Given a thermodynamic forcing, the SEQ outputs a streamfunction representing the balanced circulation induced by said heating. While this method links thermodynamics and kinematics, the physical assumptions used to derive the SEQ (hydrostatic, thermal wind balance) mean that the thermal forcing needs to be averaged over a long period of time. Additionally, the SEQ only provides the 2D solutions for azimuthally-averaged 2D thermal forcing in radius-height coordinates, which removes crucial spatial context. Finally, \citeA{bui2009balanced} noted that the balance solution substantially underestimates the boundary layer inflow, which is problematic for surface intensity assessments. While an unbalanced version of the SEQ exists \cite{ji2023does}, the extended SEQ is still a 2D diagnostic framework. Following the above discussions, our ML method complements SEQ by allowing 3D thermal forcing as an input, which enables analyses of the effect of asymmetric thermal forcing on surface winds (Section ~\ref{main:result_axisymmetric}).

Our ML framework extracts one time-invariant pattern per variable. The intensification rates predicted by the framework are linearly related to the spatial similarities between the radiative structures of individual samples and the extracted pattern. Our approach complements traditional composite analysis by removing subjectivity in the definition of composites. Finally, our framework highlights specific small-scale anomalies most relevant to information by enabling quantifiable assessments of each anomaly's contribution to ML predictions.

\subsection{Anticipating the Response of Tropical Cyclones to Radiative Perturbations} \label{main_intervention}

One of the main advantages of ML models compared to traditional physics-based models is that ML models are inexpensive to run once they are trained. We may leverage this characteristic of ML models for scientific discovery. Specifically, we may treat ML models as an efficient hypothesis generator and a framework for simple hypothesis testing.

We design a sensitivity experiment using the trained Haiyan models to understand whether the prominent wavenumber-1 structure in $\mu_{LW}$ means that asymmetric longwave anomaly is more critical to intensification than axisymmetric ones. These questions are easily answerable by feeding the ML models with perturbed input, i.e., synthetic structures. Here, we present the full definition of axisymmetric and asymmetric anomalies used in the intervention experiments.

Taking longwave radiation as an example, its mean contribution to the ML intensification forecast is proportional to $\mu_{LW} (t)$ (Eq. \ref{eq_main:mu_LW}). We would like to show that adding an asymmetric pattern like $\Pi_{\mu LW}$ causes the ML model to predict higher intensification rates than adding an axisymmetric pattern. In the asymmetric pattern experiment, we perturb the raw data in cartesian coordinates by adding the learned pattern to the training-mean longwave cooling field: 
\begin{equation} \label{sensitivity:perturb_asym}
\overline{LW_{training}}\pm \Pi_{\mu LW},
\end{equation} 
which tests the effect of adding or removing extra radiation to specific areas in the developing TC. For the case of the axisymmetric pattern experiment, we perturb the raw data with the azimuthal mean of the extracted $\mu_{LW}$ pattern:

\begin{equation}
\overline{LW_{training}}\pm \gamma \overline{\Pi_{\mu LW}}^{\theta},
\label{sensitivity:perturb_axisym}
\end{equation} 
\noindent where $\gamma$ is a multiplication factor that ensures that both synthetic structures have the same spatial variance and $(\overline{\ }^{\theta}) $ is the azimuthal mean operator. Practically, every grid point in the axisymmetric synthetic structure is multiplied by the ratio between the standard deviation of the asymmetric and axisymmetric synthetic structure.

\begin{figure*}
    \includegraphics[width=\textwidth]{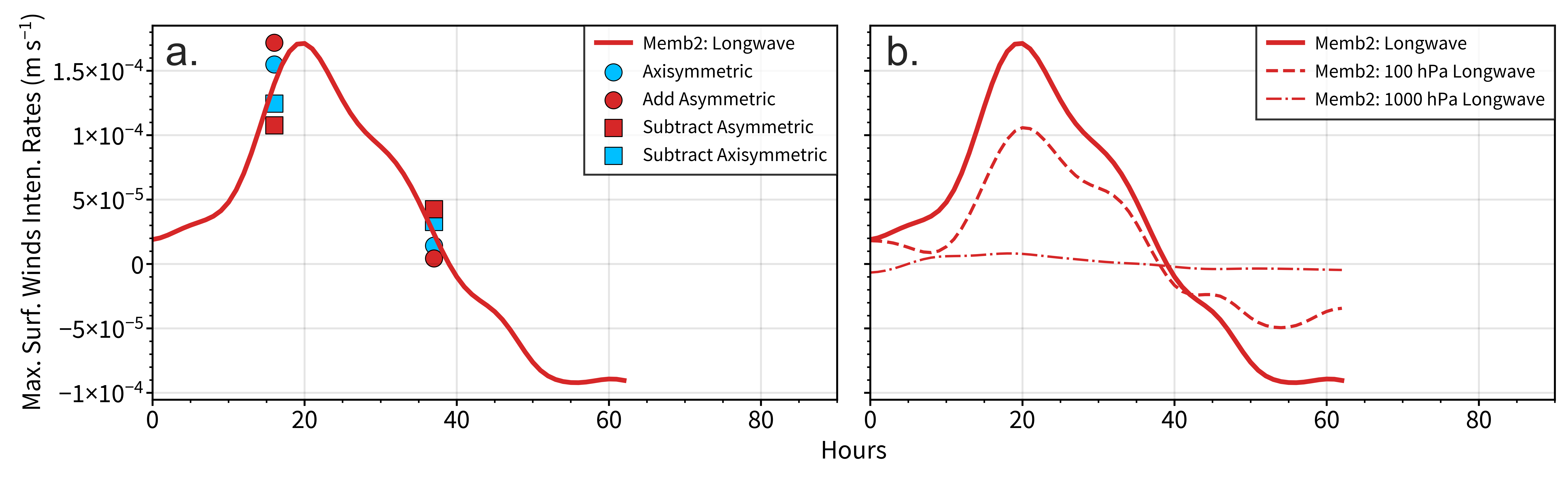}
    \caption{Providing the trained ML models with perturbed inputs gives us the first hints as to how the tropical cyclones might behave in a true intervention setting. Panel (a) shows how perturbating the Haiyan Member 2 at Hour 15 with the asymmetric $\Pi_{\mu LW}$ (red dots and squares) changes the longwave contribution to the ML prediction (red line) compared to perturbing the same sample with an axisymmetric version of $\Pi_{\mu LW}$ (blue dots and squares). Panel (b) shows that upper-level anomalies (100 hPa; dashed red) contributes more to the overall longwave contribution (red line) than the lower tropospheric anomalies (1000 hPa; dashed dotted red).}
    \label{fig:5:sensitivity}
\end{figure*}

Feeding the VED with the new inputs perturbed by the synthetic asymmetric structure predicts higher longwave contributions to intensification during the early intensification stage. Conversely, the axisymmetric synthetic structure leads to a smaller response in the VED prediction (Fig.~\ref{fig:5:sensitivity}a). VED predictions are only weakly sensitive to the synthetic perturbations later in the TC's life cycle. These results broadly agree with our conjecture that asymmetric longwave forcing leads to faster early TC intensification in the data-driven models.

\section{Conclusion}
Recent evidence \cite<e.g., >{Ruppert_etal2020} highlights the significance of radiation in early TC development. However, the precise role of radiation in realistic TCs and whether asymmetric structures in radiation impact TCs differently from axisymmetric ones has remained underexplored. We fill this knowledge gap by employing a data-driven, interpretable, stochastic, machine learning model (VED) on convective-permitting simulations of two TCs to estimate the transfer function between 3D radiative patterns and surface wind intensification. We optimize the model architecture to enhance the likelihood of extracting physically meaningful patterns. This optimization involves (i) maximizing surface intensification prediction skills by optimally combining different asymmetric radiative patterns and (ii) regressing to the unconditional distribution of intensification rates when radiation is uninformative. In both case studies, our model finds spatially coherent, physically meaningful radiation patterns from complex, high-dimensional simulation output data.

The VED architecture allows the quantification of uncertainties in both the model predictions and the extracted radiation patterns. We provide several examples where we leverage this information for scientific discovery. By intentionally making the model learn exclusively from radiation, the errors and uncertainties in ML predictions are quantifiable indicators of how relevant radiation is to TC intensification. Our findings reveal that longwave radiation exerts a larger influence on intensification than shortwave. The radiation is mostly only relevant in the early genesis phase (Section ~\ref{main:result_timeseries}), consistent with the results from prior physical modeling experiments. 

The ML model extracts three-dimensional structures in radiation and quantifies the effect of different radiative anomalies on intensification. Our results suggest that the combination of downshear left deep convective development in the inner core and low tropospheric shallow clouds in the downshear right quadrant may be used as valuable predictors for the rate of intensification during the TC genesis phase (Section ~\ref{main:result_asymmetric}). The deep convective and shallow cloud signals have a wavenumber-1 spatial asymmetry. Our VED model's linearity is useful here because we can, e.g., decompose by vertical level to conclude that the upper-level radiative signal likely impacts the surface intensification the most. The VED-extracted three-dimensional longwave anomaly structure for Haiyan has a strong spatial correlation with the broader surface wind intensification in the northern quadrants, which is potentially helpful to highlight where longwave radiative anomalies should be distributed that can potentially yield the most impact on TC intensification.

We believe that the strength of the ML framework here lies in its simplicity. The entirely linear nature of the model ensures full interpretability and decomposability, providing clarity on how the model extracts structures from WRF outputs. Our case studies demonstrate the value of sacrificing some model performance for interpretability and how interpretability leads to scientific discovery. The simple linear VED models also show superior skills to the more complex ML models in low sample size regimes (Section ~\ref{subsection: choosing}). Our approach is a valuable addition to existing data-driven approaches for scientific exploration and is particularly useful in situations with limited training data. The architecture is flexible, allowing for an easy introduction of nonlinearity and making it adaptable to other prediction problems in more complex systems.

Looking ahead, interpretable ML architectures can potentially be used to (i) reveal the source of forecast spreads in ensemble model predictions for extreme winds, e.g., by identifying the dominant uncertainty-adding circulation patterns, and (ii) quantifying the impact of different physical variables on uncertainties in climate models. We can also adopt the framework to forecast the spin-up of local winds in different shear-relative quadrants for a clearer understanding of the role of radiation in the genesis of TCs. Finally, a logical next step to build trust in the ML model explanations is to use the learned structures for targeted sensitivity experiments with physics-based numerical models, which should clarify the causal relationship between these structures and TC intensification.

\section{Open Research}
The code used to train the neural networks and to produce all figures of this manuscript is hosted on Github (\url{https://github.com/freddy0218/2024_TCG_VED}) and stored in a DOI-assigned public repository \cite{Tam_Zeonodo_TCG_VED24_github}. The processed PC time series data and trained models are archived on a separate, DOI-assigned public repository \cite{Tam_Zeonodo_TCG_VED24}. The post-processed longwave and shortwave WRF radiation fields for the two Haiyan ensemble members are also included in the archive for recreating the results shown in Figure \ref{fig:prediction_uncertainty} and \ref{fig:latent_structures}. All WRF namelist settings files to recreate the WRF simulations used in this study are archived in a separate repository \cite{Ruppert_WRF_zenodo} with postprocessing scripts. The PyTorch \cite{paszke2019pytorch} framework is used to train all the machine learning models in this manuscript. The binary files of PyTorch are available for installation via the Anaconda platform. The Optuna optimization tool can be accessed on \url{https://github.com/optuna/optuna}. We provide a short Jupyter tutorial with the minimal steps required to use the trained models to get the intensification rate predictions and the extracted mean longwave structure. This file can be assessed from the repository \cite{Tam_Zeonodo_TCG_VED24} ($\mathrm{minimal_{} example.ipynb}$).

\acknowledgments
This research was supported by the canton of Vaud in Switzerland. J.H.R. acknowledges funding support from the National Science Foundation under grants AGS 1712290 and 2331120. The authors acknowledge the Scientific Computing and Research Support Unit (DCSR) and Margot Sirdey at the University of Lausanne for providing the necessary computational resources and technical support. The authors also acknowledge Saranya Ganesh S., Milton Gomez, Louis Poulain-Auzéau, Allison A. Wing, and Caroline Muller for fruitful discussions at different stages of this manuscript's conceptualization.

%
%
%
%
\appendix
\section{Derivation steps for the Effective Weights, Biases, and Constants} \label{appendix_MATH}
This appendix presents the relevant steps to derive the Effective Weights and bias of the overall VED model (Sections \ref{math:2}, \ref{math:3}), followed by the scaling factors and logarithmic variance constants.

We first focus on the model bias term. The decoding (prediction) module of the VED model can be mathematically expressed by: 
\begin{equation} \label{MLeq:layer2}
\begin{aligned}\left(\frac{dV_{surf}}{dt}\right)_{24\mathrm{hr}} & =a_{2,LW}{\cal N}\left(\mu_{LW},e^{\log\sigma_{LW}^{2}}\right)\\
 & +a_{2,SW}{\cal N}\left(\mu_{SW},e^{\log\sigma_{SW}^{2}}\right)\\
 & +b_{2}.
\end{aligned}
\end{equation}
We expand \ref{MLeq:layer2} with \ref{normalization:PC} and \ref{MLeq:layer1}, the mean structure scalars ($\mu_{LW}$,$\mu_{SW}$) in \ref{MLeq:layer2} becomes:
\begin{equation} \label{MLeq:bias_expand}
    \begin{aligned}\mu_{LW} & =a_{2,LW} \left(b_{1,LW,\mu} + \sum_{i=1}^{n_{LW}}\frac{a_{1,LW,\mu,i}}{\sigma\left(PC_{i,LW}\right)}PC_{i,LW}-\sum_{i=1}^{n_{LW}}\frac{a_{1,LW,\mu,i}}{\sigma\left(PC_{i,LW}\right)}\overline{PC_{i,LW}}\right),\\
    \mu_{SW} & =a_{2,SW} \left(b_{1,SW,\mu} + \sum_{i=1}^{n_{SW}}\frac{a_{1,SW,\mu,i}}{\sigma\left(PC_{i,SW}\right)}PC_{i,SW}-\sum_{i=1}^{n_{SW}}\frac{a_{1,SW,\mu,i}}{\sigma\left(PC_{i,SW}\right)}\overline{PC_{i,SW}}\right).
\end{aligned}
\end{equation}
The overall bias ($b$) of the VED model is the sum of the first and third terms in Eqs. \ref{MLeq:bias_expand}.

Comparing Equations \ref{fulleq:goal}, \ref{MLeq:layer2}, and \ref{MLeq:bias_expand} shows that the non-constant terms (second on the right-hand side) in Equation \ref{MLeq:bias_expand} is equivalent to the projection of radiative anomaly onto the learned patterns (Eqs. \ref{define:innerPCs}). Using longwave mean structure as an example, we have:
\begin{equation} \label{equivalence:MEAN_1}
PC_{i,\Pi_{LW, \mu}} = \lambda \frac{a_{2,LW} a_{1,LW,\mu,i}}{\sigma(PC_{i,LW})},
\end{equation}
where $\lambda$ is the proportionality coefficient for the longwave mean structure. We now assume the norm of the data-driven structure to be 1:
\begin{equation}
    1=\left\langle \Pi_{LW, \mu}\ |\ \Pi_{\mu LW}\right\rangle_{LW} =\sum_{i=1}^{n_{LW}}PC_{i,\Pi_{\mu LW}}^{2}=\lambda^{2}a_{2,LW}^{2}\sum_{i=1}^{n_{LW}}\frac{a_{1,LW,\mu,i}^{2}}{\sigma\left(PC_{i,LW}\right)^{2}},
\label{equivalence:MEAN_norm1}
\end{equation}

From Equation~\ref{equivalence:MEAN_norm1}, the $\lambda$ for longwave mean structure is:
\begin{equation} \label{equivalence:MEAN_coeff}
\lambda_{LW,\mu}=\left[\left|a_{2,LW}\right|\sqrt{\sum_{i=1}^{n_{LW}}\frac{a_{1,LW,\mu,i}^{2}}{\sigma\left(PC_{i,LW}\right)^{2}}}\right]^{-1}.
\end{equation}

The other three $\lambda$ in the VED model can be obtained with a similar procedure.

We can rewrite Equation~\ref{MLeq:layer2} with the definition of the mean structure scalars (Eqs.~\ref{MLeq:bias_expand}), Equations~\ref{MLeq:layer1}, and $\lambda$ (Eq.~\ref{equivalence:MEAN_coeff},

\begin{equation} \label{fulleq:mean_expand1}
\begin{aligned}
\left(\frac{dV_{surf}}{dt}\right)_{24\mathrm{hr}} & ={\cal N}\left({\frac{a_{2,LW}}{\lambda_{LW, \mu}}}\underbrace{\sum_{i=1}^{n_{LW}}PC_{i,\Pi\mu LW}PC_{i,LW}}_{\left\langle LW|\Pi_{\mu LW}\right\rangle }\ ,\ \sqrt{a_{2,LW}}e^{\log\sigma_{LW}^{2}}\right)\\
 & +{\cal N}\left(\frac{a_{2,SW}}{\lambda_{SW,\mu}}\underbrace{\sum_{i=1}^{n_{SW}}PC_{i,\Pi\mu SW}PC_{i,SW}}_{\left\langle SW|\Pi_{\mu SW}\right\rangle }\ ,\ \sqrt{a_{2,SW}}e^{\log\sigma_{SW}^{2}}\right)\\
 & +b
\end{aligned}
\end{equation}

To see how Equation \ref{fulleq:mean_expand1} related to Equation (\ref{fulleq:goal}), we factor out all constant terms from the normal distribution:
\begin{equation} \label{fulleq:mean_expand_factorcnst}
\begin{aligned}
\left(\frac{dV_{surf}}{dt}\right)_{24\mathrm{hr}} & =\underbrace{\frac{a_{2,LW}}{\lambda_{LW,\mu}}}_{a_{LW}}{\cal N}\left(\left\langle LW|\Pi_{\mu LW}\right\rangle_{LW} \ ,\ \sqrt{\lambda_{LW,\mu} \left|a_{2,LW}\right|} e^{\log\sigma_{LW}^{2}}\right)\\
 & +\underbrace{\frac{a_{2,SW}}{\lambda_{SW,\mu}}}_{a_{SW}}{\cal N}\left(\left\langle SW|\Pi_{\mu SW}\right\rangle_{SW} \ ,\ \sqrt{\lambda_{SW,\mu} \left|a_{2,SW}\right|} e^{\log\sigma_{SW}^{2}} \right)\\
 & +b,
\end{aligned}
\end{equation}
allowing us to identify the ``effective weights'' $a_{LW}$ and $a_{SW}$. 

We now use the definition of the logarithmic variance in the model's first (projection) layer (\ref{MLeq:layer1}) to expand the logarithmic variance term of the normal distributions:
\begin{equation} \label{fulleq:logvar_expand1}
\sqrt{\lambda_{LW,\mu} \left|a_{2,LW}\right|} \exp \left(\log\sigma_{LW}^{2}\right) = \sqrt{\lambda_{LW,\mu} \left|a_{2,LW}\right|} \exp \left(b_{1,LW,\log\sigma^{2}}+\sum_{i=1}^{n_{LW}}a_{1,LW,\log\sigma^{2},i}\times\widetilde{PC}_{i,LW}\right)
\end{equation}

\begin{equation} \label{fulleq:logvar_expand1_SW}
\sqrt{\lambda_{SW,\mu} \left|a_{2,SW}\right|} \exp\left(\log\sigma_{SW}^{2}\right) = \sqrt{\lambda_{SW,\mu} \left|a_{2,SW}\right|} \exp\left(b_{1,SW,\log\sigma^{2}}+\sum_{i=1}^{n_{SW}}a_{1,SW,\log\sigma^{2},i}\times\widetilde{PC}_{i,SW}\right)
\end{equation}

We notice that parts of Equation \ref{fulleq:logvar_expand1} and \ref{fulleq:logvar_expand1_SW} are constant, which means that we can simplify this equation further by defining variance prefactors for longwave radiation ($c_{LW}$; Eq. \ref{constant:c_lw}) and shortwave radiation ($c_{SW}$; Eq. \ref{constant:c_sw}).

\begin{equation} \label{Equation:Clw}
c_{LW} = \sqrt{\lambda_{LW,\mu} \left|a_{2,LW}\right|} \exp\left(b_{1,LW,\log\sigma^{2}}-\sum_{i=1}^{n_{LW}}a_{1,LW,\log\sigma^{2},i}\frac{\overline{PC_{i,LW}}}{\sigma\left(PC_{i,LW}\right)}\right),
\end{equation}

\begin{equation} \label{Equation:Csw}
c_{SW} = \sqrt{\lambda_{SW,\mu} \left|a_{2,SW}\right|} \exp\left(b_{1,SW,\log\sigma^{2}}-\sum_{i=1}^{n_{SW}}a_{1,SW,\log\sigma^{2},i}\frac{\overline{PC_{i,SW}}}{\sigma\left(PC_{i,SW}\right)}\right),
\end{equation}

Using the same reasoning as for the mean longwave structure (Eq.~\ref{equivalence:MEAN_coeff}), proportionality coefficient for the longwave logarithmic variance structure ($\lambda_{LW,\mathrm{log}\sigma^2}$) for the PC loadings of the logarithmic variance structure:
\begin{equation} \label{Equation:LOGVAR_coeff}
PC_{i,\Pi\log\sigma_{LW}^{2}}=\lambda_{LW,\mathrm{log}\sigma^2} \frac{a_{1,LW,\log\sigma^{2},i}}{\sigma\left(PC_{i,LW}\right)}\ \Rightarrow\ \lambda_{LW,\mathrm{log}\sigma^2}=\left(\sum_{i=1}^{n_{LW}}\frac{a_{1,LW,\log\sigma^{2},i}^{2}}{\sigma\left(PC_{i,LW}\right)^{2}}\right)^{-1/2}.
\end{equation}

\begin{equation} \label{Equation:LOGVAR_coeff_SW}
PC_{i,\Pi\log\sigma_{SW}^{2}}=\lambda_{LW,\mathrm{log}\sigma^2} \frac{a_{1,SW,\log\sigma^{2},i}}{\sigma\left(PC_{i,SW}\right)}\ \Rightarrow\ \lambda_{SW,\mathrm{log}\sigma^2}=\left(\sum_{i=1}^{n_{SW}}\frac{a_{1,SW,\log\sigma^{2},i}^{2}}{\sigma\left(PC_{i,SW}\right)^{2}}\right)^{-1/2}.
\end{equation}


For consistency with the mean structures, we transform the logarithmic variance term (Eq. \ref{fulleq:logvar_expand1}-\ref{fulleq:logvar_expand1_SW}) using Eqs~\ref{fulleq:logvar_expand1},~\ref{fulleq:logvar_expand1_SW},~\ref{Equation:LOGVAR_coeff}, and~\ref{Equation:LOGVAR_coeff_SW}, which yields:
\begin{equation} \label{logvarterm:gettingD}
c_{LW}\exp\left(\sum_{i=1}^{n_{LW}}\frac{a_{1,LW,\log\sigma^{2},i}\times PC_{i,LW}}{\sigma\left(PC_{i,LW}\right)}\right)=c_{LW}\exp\left(\underbrace{\frac{1}{\lambda_{LW,\mathrm{log}\sigma^2}}}_{d_{LW}}\times\left\langle LW\ |\ \Pi_{\log\sigma_{LW}^{2}}\right\rangle_{LW} \right)
\end{equation}

\begin{equation} \label{logvarterm:gettingD_SW}
c_{SW}\exp\left(\sum_{i=1}^{n_{SW}}\frac{a_{1,SW,\log\sigma^{2},i}\times PC_{i,SW}}{\sigma\left(PC_{i,SW}\right)}\right)=c_{SW}\exp\left(\underbrace{\frac{1}{\lambda_{SW,\mathrm{log}\sigma^2}}}_{d_{SW}}\times\left\langle SW\ |\ \Pi_{\log\sigma_{SW}^{2}}\right\rangle_{SW} \right)
\end{equation}

\section{Baseline: Principal Component Regression} \label{SI:PCRegression}
\begin{figure}
    \includegraphics[width=\textwidth]{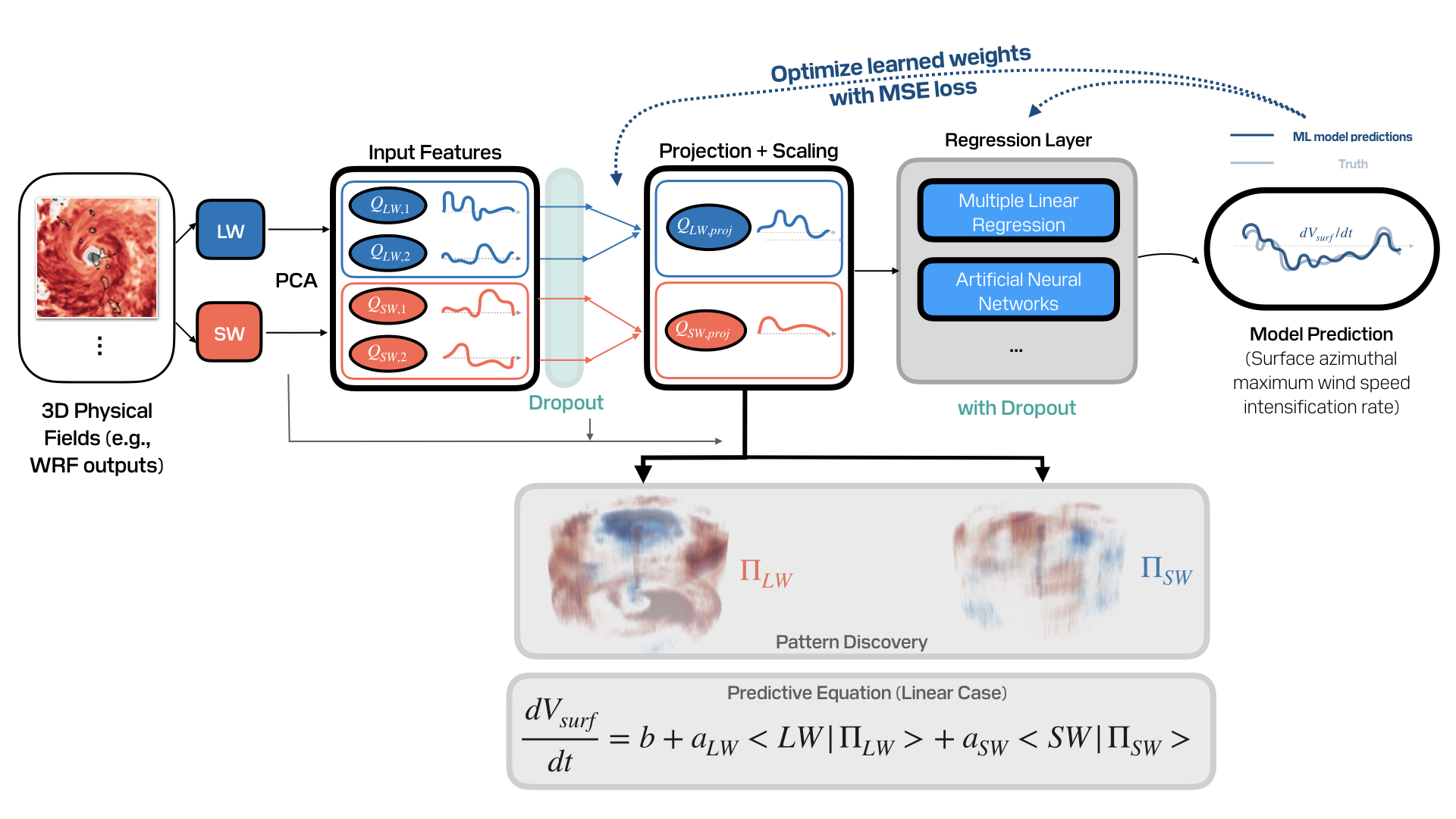}
    \caption{The interpretable linear model baseline to compare to the VED model. This baseline model introduces uncertainty with the dropout operation.}
    \label{fig:SIschematic}
\end{figure}

We implement a simple two-layer linear regression model as a baseline to examine if the more complex VED architecture improves the overall prediction skills and provides a more well-calibrated prediction uncertainty. This baseline is analogous to a two-branch principal component regression (schematic diagram in Fig. \ref{fig:SIschematic}). The baseline model only extracts two structures from the longwave and shortwave radiation information. These structures are analogous to the $\mu_{LW}$ and $\mu_{SW}$ in the VED model. Rather than extracting the uncertainty structures and sampling them with the reparameterization trick, we introduce uncertainty in the baseline model using a dropout mechanism. The dropout mechanism zeroes out a random selection of input features, allowing the model to have uncertainties in both the structure layer level and the final prediction outputs. The amount of input features this operation drops is determined by a tunable $\mathbf{dropout\ rate}$ hyperparameter that ranges from 0 to 1.

\section{Decomposition by Vertical Levels} \label{appendix_VL}
We isolate the effect of heating anomalies at different vertical levels by setting all grid point values in the input data to zero, \textit{except for those at the vertical level(s) of interest}.

For instance, if we are interested in the individual contributions of longwave radiative heating at 1000 hPa and 100 hPa, we can separate the LW field into three distinct terms:
\begin{equation} \label{sensitivity:LWdecomp}
LW = LW_{100} + LW_{1000} + LW_{200-900},
\end{equation}
where $LW_{100}$ ($LW_{1000}$) is the longwave field where all grid points except those at 100 (1000) hPa are zeroed out, $LW_{200-900}$ is the longwave field where grid points at 100 and 1000 hPa are zeroed out.

Projecting the three components of the right-hand side of Eq.~\ref{sensitivity:LWdecomp} onto the longwave PC eigenvectors yields the same decomposition for the PC loadings:
\begin{equation} \label{sensitivity:PCdecomp}
\forall i \in \llbracket 1,n_{LW} \rrbracket \ , \  PC_{i,LW} = PC_{i,LW_{100}} + PC_{i,LW_{1000}} + PC_{i,LW_{200-900}}
\end{equation}

After standardization, Equation (\ref{sensitivity:PCdecomp}) can be separated into a constant part: 

\begin{equation} \label{sensitivity:PCstandardization_const}
\underbrace{\left(\frac{\overline{PC_{i,LW_{100}}}+\overline{PC_{i,LW_{1000}}}+\overline{PC_{i,LW_{200-900}}}}{\sigma(PC_{i,LW})}\right)}_{\overline{PC_{i,LW}}}
\end{equation}

and a part that varies in time:
\begin{equation} \label{sensitivity:PCstandardization_PC}
\underbrace{\left[\frac{PC_{i,LW_{100}}}{\sigma(PC_{i,LW})} +  \frac{PC_{i,LW_{1000}}}{\sigma(PC_{i,LW})} + \frac{PC_{i,LW_{900-200}}}{\sigma(PC_{i,LW})}\right]}_{PC_{i,LW}},
\end{equation}

\begin{figure*}
    \includegraphics[width=\textwidth]{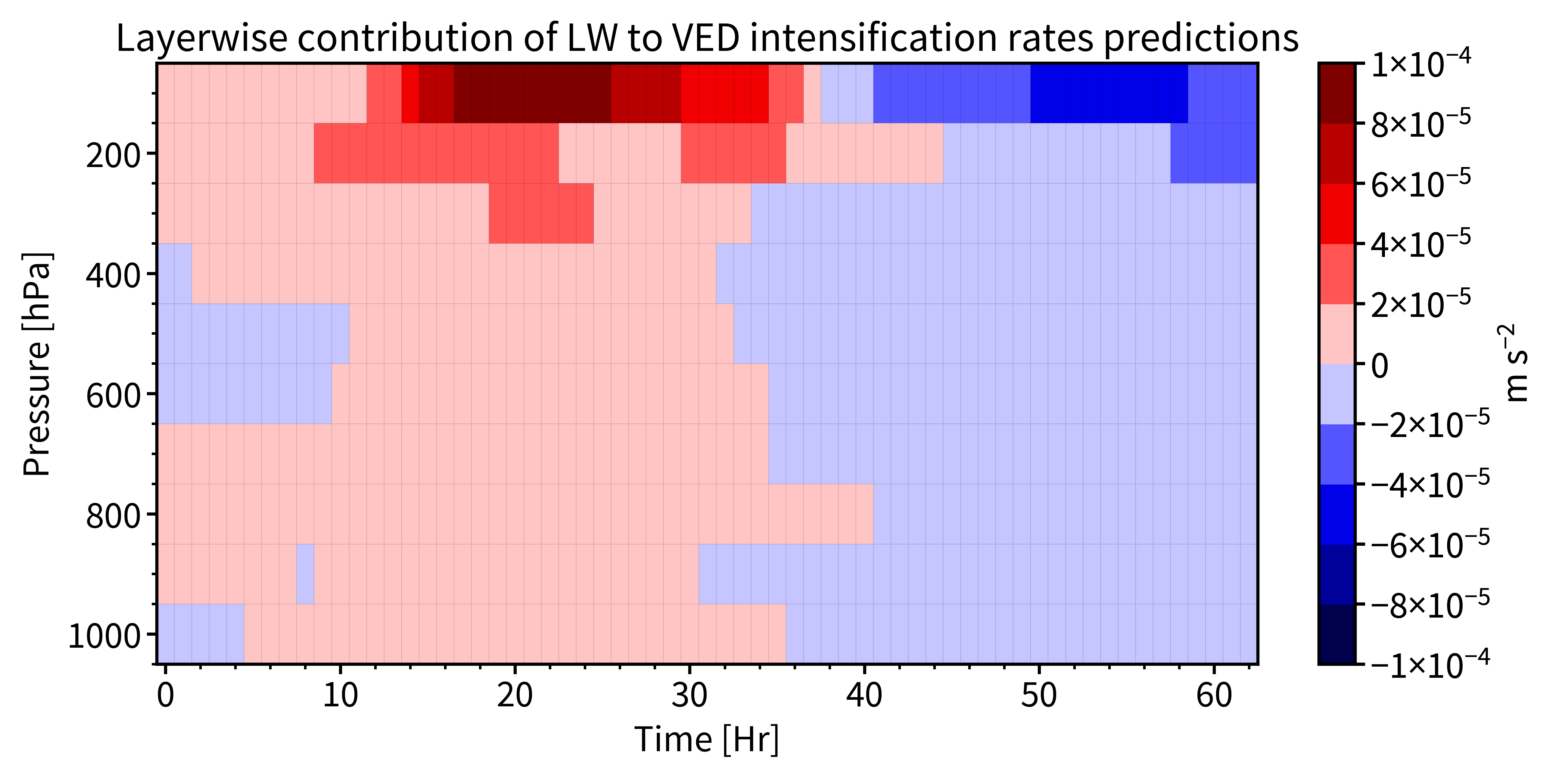}
    \caption{Decomposing the contribution of longwave heating to VED intensity change predictions reveals the important contribution of upper-level radiative anomalies. To create this Hovmöller diagram, we conducted 10 experiments, where all anomalies are zeroed out except at a specific vertical level on a coarse 10-level grid, spaced at 100 hPa intervals. This approach quantifies the contribution to changes in VED TC intensification predictions in acceleration units (m/$s^2$).}
    \label{fig:SI:layer}
\end{figure*}

Substituting Equation \ref{sensitivity:PCstandardization_PC} into Equation \ref{fulleq:mean_expand1} and expanding the first term in Equation \ref{fulleq:mean_expand1}, we can decompose surface intensification into additive terms that match the longwave radiative heating field's decomposition of Eq. \ref{sensitivity:LWdecomp}.

\begin{equation} \label{sensitivity:define}
\begin{aligned}
\frac{dV_{surf, 24}}{dt} = &\left(\frac{dV_{LW, 100 hPa}}{dt}+\frac{dV_{LW, 200-1000 hPa}}{dt}\right) \\
&+ \frac{dV_{SW, 100-1000 hPa}}{dt}+b,
\end{aligned}
\end{equation}
where $\frac{dV_{LW, 100 hPa}}{dt}$ quantifies how radiation anomalies at 100 hPa affect the surface intensification. Figure \ref{fig:SI:layer} shows an example of how we use this decomposition technique to illustrate the contributions of longwave radiative anomalies at different vertical levels to Haiyan Member 2.

\bibliography{agusample}

%
%
%
%
%

\end{document}


%
%


\title{Supporting Information for ``Identifying Three-Dimensional Radiative Patterns Associated with Early Tropical Cyclone Intensification''}
%
%

%
%



\authors{Frederick Iat-Hin Tam\affil{1,2}, Tom Beucler\affil{1,2}, James H. Ruppert Jr.\affil{3}}

\affiliation{1}{Faculty of Geosciences and Environment, University of Lausanne, Switzerland}
\affiliation{2}{Expertise Center for Climate Extremes, University of Lausanne, Switzerland}
\affiliation{3}{School of Meteorology, University of Oklahoma, Norman, OK, USA}

%
%

%

\begin{article}

%
%



\section{Details of the Machine Learning Framework} 
The proposed interpretable machine learning model consists of three parts: (i) a dimensionality reduction layer to process full complexity input fields in physical units, (ii) a linear encoder that condenses all low-dimensionality representations of a physical field into a projection that is most useful to the prediction task, and (iii) a final linear decoder that uses multiple linear regression to give the 24-hour TC intensification rates based on a given set of input fields.

\subsection{Mathematical Construction} \label{Details:Maths}
This section presents full mathematical descriptions of the three different parts of the ML framework. 

We choose to use Principal Component Analysis (PCA) to preserve linearity in the dimensionality reduction layer. PCA linearly transforms input physical fields ($X_{i} $) into combinations of orthogonal eigenvectors (PCA modes; $\Pi_{X_{i}}(z,r,\theta)$) and their corresponding time evolution (PC loadings time series; $PC_{X_{i}}(t)$):
\begin{equation} \label{define:PC}
    X_{i}\left(t,z,r,\theta\right)-\overline{X_{i}\left(t,z,r,\theta\right)} = \sum_{i=1}^{N} PC_{X_{i}}(t) \Pi_{X_{i}}(z,r,\theta),
\end{equation}
where $N$ is the number of retained PCA modes. Different loadings $PC_{X_{i}}(t)$ given by the PCA will be used as inputs to the ML model, instead of the raw field $X_{i}$ in its full spatiotemporal complexity.\\

Following best machine learning practices, all $PC_{X_{i}}(t)$ time series are standardized to have a mean of 0 and a variance of 1 to avoid high variance in the regression weights:
\begin{equation} \label{normalization:PC}
    \widetilde{PC}_{X_i} (t) = \frac{PC_{X_{i}}(t)-\overline{PC_{X_{i}}(t)}}{\sigma_{PC_{X_{i}}(t)}},
\end{equation}
where $\overline{PC_{X_{i}}}$ and $\sigma_{PC_{X_{i}}(t)}$ are the mean and standard deviation of the PC loadings, calculated over the training set.

In the encoder module of our VED model, we combine different PCA modes for longwave radiation and shortwave radiation into two ``mean structures'' ($\Pi_{LW,\mu},\Pi_{SW,\mu}$) and two ``logarithmic variance structures'' ($\Pi_{LW, log\sigma^2},\Pi_{SW,log\sigma^2} $). These data-driven structures do not vary in time. The model's first layer involves projecting the original radiation fields onto the learned mean and logarithmic variance structures. We now introduce an inner product that will help us define this projection:
\begin{equation} \label{define:innerproduct}
\left\langle PC_{X_1}|PC_{X_2}\right\rangle_{X_3} = \sum_{i=1}^{N} PC_{X_1}^i PC_{X_2}^i,
\end{equation}
the $X_3$ indicates that the inner product and the projection are defined with respect to the variable for which the PC decomposition and orthogonal modes are calculated.

Using this notation, the 24-hr surface intensity intensification rates may be computed with the projections obtained in the first model layer:
\begin{equation} \label{fulleq:goal}
\begin{aligned}
\left(\frac{dV_{surf}}{dt}\right)_{24\mathrm{hr}} & =a_{LW}{\cal N}\left(\left\langle LW|\Pi_{\mu LW}\right\rangle_{LW} \ ,\ c_{LW}e^{d_{LW}\left\langle LW\ |\ \Pi_{\log\sigma_{LW}^{2}}\right\rangle_{LW} }\right)\\
 & +a_{SW}{\cal N}\left(\left\langle SW|\Pi_{\mu SW}\right\rangle_{SW} \ ,\ c_{SW}e^{d_{SW}\left\langle LW\ |\ \Pi_{\log\sigma_{SW}^{2}}\right\rangle_{SW} }\right)\\
 & +b,
\end{aligned}
\end{equation}
where the input of the final linear decoder is given by randomly sampling the normal distributions constructed with the mean and logarithmic variances calculated by the linear encoder for longwave and shortwave radiation.
We will show that the ``effective weight'' to calculate the contribution of longwave radiation (in physical units) to TC intensification can be expressed by:
\begin{equation} \label{effectweight:LW}
a_{LW} = \left|a_{2,LW}\right|\sqrt{\sum_{i=1}^{n_{LW}}\frac{a_{1,LW,\mu,i}^{2}}{\sigma\left(PC_{i,LW}\right)^{2}}},
\end{equation}
whereas the weight of the shortwave contribution is
\begin{equation} \label{effectweight:SW}
a_{SW} = \left|a_{2,SW}\right|\sqrt{\sum_{i=1}^{n_{SW}}\frac{a_{1,SW,\mu,i}^{2}}{\sigma\left(PC_{i,SW}\right)^{2}}}.
\end{equation}
We will also show that the overall model bias ($b$) is given by the following equation:
\begin{equation} \label{bias}
\begin{aligned}
b =&b_2 + a_{2,LW}\left(b_{1,LW,\mu} - \sum_{i=1}^{n_{LW}} \frac{a_{1,LW,\mu,i} \overline{PC_{i,LW}}}{\sigma(PC_{i,LW})}\right) \\
&+a_{2,SW}\left(b_{1,SW,\mu} - \sum_{i=1}^{n_{SW}} \frac{a_{1,SW,\mu,i} \overline{PC_{i,SW}}}{\sigma(PC_{i,SW})}\right),
\end{aligned}
\end{equation}
and that the four constants involved in the variance calculation ($c_{LW}, c_{SW}, d_{LW}, d_{SW}$) are given by the following equations:
\begin{equation} \label{constant:c_lw}
c_{LW}=\left(\sum_{i=1}^{n_{LW}}\frac{a_{1,LW,\mu,i}^{2}}{\sigma\left(PC_{i,LW}\right)^{2}}\right)^{-1/4}\exp\left(b_{1,LW,\log\sigma^{2}}-\sum_{i=1}^{n_{LW}}a_{1,LW,\log\sigma^{2},i}\frac{\overline{PC_{i,LW}}}{\sigma\left(PC_{i,LW}\right)}\right),
\end{equation}

\begin{equation} \label{constant:c_sw}
c_{SW}=\left(\sum_{i=1}^{n_{SW}}\frac{a_{1,SW,\mu,i}^{2}}{\sigma\left(PC_{i,SW}\right)^{2}}\right)^{-1/4}\exp\left(b_{1,SW,\log\sigma^{2}}-\sum_{i=1}^{n_{SW}}a_{1,SW,\log\sigma^{2},i}\frac{\overline{PC_{i,SW}}}{\sigma\left(PC_{i,SW}\right)}\right),
\end{equation}

\begin{equation} \label{constant:d_lw}
d_{LW}=\sqrt{\sum_{i=1}^{n_{LW}}\frac{a_{1,LW,\log\sigma^{2},i}^{2}}{\sigma\left(PC_{i,LW}\right)^{2}}},
\end{equation}

\begin{equation} \label{constant:d_sw}
d_{SW}=\sqrt{\sum_{i=1}^{n_{SW}}\frac{a_{1,SW,\log\sigma^{2},i}^{2}}{\sigma\left(PC_{i,SW}\right)^{2}}}.
\end{equation}

The parameters $b_2$, $a_{2,LW}$, and $a_{2,SW}$ represent the bias term, weights for longwave radiation, and the weights for shortwave radiation in the final regression layer, respectively. The constants also includes the weights and biases from the longwave mean structure encoder ($\boldsymbol{a_{1,LW,\mu}}$, $\boldsymbol{b_{1,LW,\mu}}$), the shortwave mean structure encoder ($\boldsymbol{a_{1,SW,\mu}}$, $\boldsymbol{b_{1,SW,\mu}}$), the temporal average of PC time series ($\overline{PC_{i,LW}}$, $\overline{PC_{i,SW}}$), and their standard deviations ($\sigma(PC_{i,LW})$, $\sigma(PC_{i,SW})$). Using this notation, the four outputs of our model's first (projection) layer are given by the following four equations:
\begin{equation} \label{MLeq:layer1}
    \begin{aligned}\mu_{LW} & =b_{1,LW,\mu}+\sum_{i=1}^{n_{LW}}a_{1,LW,\mu,i}\times\widetilde{PC}_{i,LW},\\
\log\sigma_{LW}^{2} & =b_{1,LW,\log\sigma^{2}}+\sum_{i=1}^{n_{LW}}a_{1,LW,\log\sigma^{2},i}\times\widetilde{PC}_{i,LW},\\
\mu_{SW} & =b_{1,SW,\mu}+\sum_{i=1}^{n_{SW}}a_{1,SW,\mu,i}\times\widetilde{PC}_{i,SW},\\
\log\sigma_{SW}^{2} & =b_{1,SW,\log\sigma^{2}}+\sum_{i=1}^{n_{SW}}a_{1,SW,\log\sigma^{2},i}\times\widetilde{PC}_{i,SW},
\end{aligned}
\end{equation}
and the output of our model's second (prediction) layer is given by the following equation:
\begin{equation} \label{MLeq:layer2}
\begin{aligned}\left(\frac{dV_{surf}}{dt}\right)_{24\mathrm{hr}} & =a_{2,LW}{\cal N}\left(\mu_{LW},e^{\log\sigma_{LW}^{2}}\right)\\
 & +a_{2,SW}{\cal N}\left(\mu_{SW},e^{\log\sigma_{SW}^{2}}\right)\\
 & +b_{2}.
\end{aligned}
\end{equation}

Combining equations \ref{normalization:PC}, \ref{MLeq:layer1}, and \ref{MLeq:layer2} yields:
\begin{equation}
\begin{aligned}\left(\frac{dV_{surf}}{dt}\right)_{24\mathrm{hr}} & =a_{2,LW}{\cal N}\left(\sum_{i=1}^{n_{LW}}\frac{a_{1,LW,\mu,i}}{\sigma\left(PC_{i,LW}\right)}PC_{i,LW},e^{\log\sigma_{LW}^{2}}\right)\\
 & +a_{2,SW}{\cal N}\left(\sum_{i=1}^{n_{LW}}\frac{a_{1,SW,\mu,i}}{\sigma\left(PC_{i,SW}\right)}PC_{i,SW},e^{\log\sigma_{SW}^{2}}\right)\\
 & +b.
\end{aligned}
\label{fulleq:insertnormalization}    
\end{equation}

We can rewrite Equation \ref{fulleq:insertnormalization} to include the prefactors $a_{2,LW}$ and $a_{2,SW}$ in the mean and variances of the two normal distributions used to calculate the intensification rate:
\begin{equation} \label{fulleq:a2_inside_N}
\begin{aligned}\left(\frac{dV_{surf}}{dt}\right)_{24\mathrm{hr}} & ={\cal N}\left(\sum_{i=1}^{n_{LW}}\frac{a_{2,LW}a_{1,LW,\mu,i}}{\sigma\left(PC_{i,LW}\right)}PC_{i,LW},\sqrt{a_{2,LW}} e^{\log\sigma_{LW}^{2}}\right)\\
 & +{\cal N}\left(\sum_{i=1}^{n_{LW}}\frac{a_{2,SW}a_{1,SW,\mu,i}}{\sigma\left(PC_{i,SW}\right)}PC_{i,SW},\sqrt{a_{2,SW}}e^{\log\sigma_{SW}^{2}}\right)\\
 & +b.
\end{aligned}    
\end{equation}


We can now use the definition of the inner product (Eq \ref{define:innerproduct}) to interpret our model's first layer as a linear projection onto four data-driven structures: $ \left(\Pi_{\mu LW},\Pi_{\mu SW},\Pi_{\log\sigma^{2}LW},\Pi_{\log\sigma^{2}SW}\right) $. According to the inner product's definition (given by Equation \ref{define:innerproduct}), the PC loadings $\left(PC_{i,\Pi_{\mu LW}},PC_{i,\Pi_{\mu SW}},PC_{i,\Pi_{\log\sigma^{2}LW}},PC_{i,\Pi_{\log\sigma^{2}SW}}\right) $ of these four structures obey the following equations:
\begin{equation}
\begin{cases} \label{define:innerPCs}
\left\langle LW\ |\ \Pi_{\mu LW}\right\rangle_{LW}  & =\sum_{i=1}^{n_{LW}}PC_{i,LW}\times PC_{i,\Pi_{\mu LW}},\\
\left\langle LW\ |\ \Pi_{\log\sigma^{2}LW}\right\rangle_{LW}  & =\sum_{i=1}^{n_{LW}}PC_{i,LW}\times PC_{i,\Pi_{\log\sigma^{2}LW}},\\
\left\langle SW\ |\ \Pi_{\mu SW}\right\rangle_{SW}  & =\sum_{i=1}^{n_{LW}}PC_{i,SW}\times PC_{i,\Pi_{\mu SW}},\\
\left\langle SW\ |\ \Pi_{\log\sigma^{2}SW}\right\rangle_{SW}  & =\sum_{i=1}^{n_{LW}}PC_{i,LW}\times PC_{i,\Pi_{\log\sigma^{2}SW}}.
\end{cases}
\end{equation}
Substituting Equation \ref{fulleq:a2_inside_N} into Equation \ref{fulleq:goal} and building upon the equivalence of equations \ref{fulleq:goal} and \ref{fulleq:insertnormalization}, we see that the PC loadings of the longwave mean structure $PC_{i,\Pi_{\mu LW}}$ are proportional to the product of the first two layers' weights: 
\begin{equation} \label{equivalence:MEAN_1}
PC_{i,\Pi_{\mu LW}} = \lambda \frac{a_{2,LW} a_{1,LW,\mu,i}}{\sigma(PC_{i,LW})},
\end{equation}
where we have used $\lambda $ to denote the proportionality coefficient in the case of the longwave ``mean structure''. To simplify the projection's definition, the norm of the data-driven structure is assumed to be 1:
\begin{equation}
    1=\left\langle \Pi_{\mu LW}\ |\ \Pi_{\mu LW}\right\rangle_{LW} =\sum_{i=1}^{n_{LW}}PC_{i,\Pi_{\mu LW}}^{2}=\lambda^{2}a_{2,LW}^{2}\sum_{i=1}^{n_{LW}}\frac{a_{1,LW,\mu,i}^{2}}{\sigma\left(PC_{i,LW}\right)^{2}}
\label{equivalence:MEAN_norm1}
\end{equation}

This means that the proportionality coefficient $\lambda$ in Equation (\ref{equivalence:MEAN_1}) is:
\begin{equation} \label{equivalence:MEAN_coeff}
\lambda=\left[\left|a_{2,LW}\right|\sqrt{\sum_{i=1}^{n_{LW}}\frac{a_{1,LW,\mu,i}^{2}}{\sigma\left(PC_{i,LW}\right)^{2}}}\right]^{-1}.
\end{equation}


Substituting Equation (\ref{equivalence:MEAN_coeff}) into Equation (\ref{fulleq:insertnormalization}) and taking the definition of the inner products (Eq. \ref{define:innerproduct}) into account, we obtain:
\begin{equation} \label{fulleq:mean_expand1}
\begin{aligned}
\left(\frac{dV_{surf}}{dt}\right)_{24\mathrm{hr}} & ={\cal N}\left(\sqrt{\sum_{i=1}^{n_{LW}}\frac{a_{1,LW,\mu,i}^{2}}{\sigma\left(PC_{i,LW}\right)^{2}}}\underbrace{\frac{a_{2,LW}}{\mathrm{sign}\left(a_{2,LW}\right)}}_{\left|a_{2,LW}\right|}\underbrace{\sum_{i=1}^{n_{LW}}PC_{i,\Pi\mu LW}PC_{i,LW}}_{\left\langle LW|\Pi_{\mu LW}\right\rangle }\ ,\ \sqrt{a_{2,LW}}e^{\log\sigma_{LW}^{2}}\right)\\
 & +{\cal N}\left(\sqrt{\sum_{i=1}^{n_{SW}}\frac{a_{1,SW,\mu,i}^{2}}{\sigma\left(PC_{i,SW}\right)^{2}}}\underbrace{\frac{a_{2,SW}}{\mathrm{sign}\left(a_{2,SW}\right)}}_{\left|a_{2,SW}\right|}\underbrace{\sum_{i=1}^{n_{SW}}PC_{i,\Pi\mu SW}PC_{i,SW}}_{\left\langle SW|\Pi_{\mu SW}\right\rangle }\ ,\ \sqrt{a_{2,SW}}e^{\log\sigma_{SW}^{2}}\right)\\
 & +b
\end{aligned}
\end{equation}

To see how Equation \ref{fulleq:mean_expand1} related to Equation (\ref{fulleq:goal}), we factor out all constant terms from the normal distribution:
\begin{equation} \label{fulleq:mean_expand_factorcnst}
\begin{aligned}
\left(\frac{dV_{surf}}{dt}\right)_{24\mathrm{hr}} & =\underbrace{\left|a_{2,LW}\right|\sqrt{\sum_{i=1}^{n_{LW}}\frac{a_{1,LW,\mu,i}^{2}}{\sigma\left(PC_{i,LW}\right)^{2}}}}_{a_{LW}}{\cal N}\left(\left\langle LW|\Pi_{\mu LW}\right\rangle_{LW} \ ,\ \frac{e^{\log\sigma_{LW}^{2}}}{\left(\sum_{i=1}^{n_{LW}}\frac{a_{1,LW,\mu,i}^{2}}{\sigma\left(PC_{i,LW}\right)^{2}}\right)^{1/4}}\right)\\
 & +\underbrace{\left|a_{2,SW}\right|\sqrt{\sum_{i=1}^{n_{SW}}\frac{a_{1,SW,\mu,i}^{2}}{\sigma\left(PC_{i,SW}\right)^{2}}}}_{a_{SW}}{\cal N}\left(\left\langle SW|\Pi_{\mu SW}\right\rangle_{SW} \ ,\ \frac{e^{\log\sigma_{SW}^{2}}}{\left(\sum_{i=1}^{n_{SW}}\frac{a_{1,SW,\mu,i}^{2}}{\sigma\left(PC_{i,SW}\right)^{2}}\right)^{1/4}}\right)\\
 & +b,
\end{aligned}
\end{equation}
allowing us to identify the ``effective weights'' $a_{LW}$ and $a_{SW}$. We now use the definition of the logarithmic variance in the model's first (projection) layer (\ref{MLeq:layer1}) to expand the logarithmic variance term of the normal distributions:
\begin{equation} \label{fulleq:logvar_expand1}
\begin{aligned}
\left(\frac{dV_{surf}}{dt}\right)_{24\mathrm{hr}} & =a_{LW}{\cal N}\left(\left\langle LW|\Pi_{\mu LW}\right\rangle_{LW} \ ,\ \frac{b_{1,LW,\log\sigma^{2}}+\sum_{i=1}^{n_{LW}}a_{1,LW,\log\sigma^{2},i}\times\widetilde{PC}_{i,LW}}{\left(\sum_{i=1}^{n_{LW}}\frac{a_{1,LW,\mu,i}^{2}}{\sigma\left(PC_{i,LW}\right)^{2}}\right)^{1/4}}\right)\\
 & +a_{SW}{\cal N}\left(\left\langle SW|\Pi_{\mu SW}\right\rangle_{SW} \ ,\ \frac{b_{1,SW,\log\sigma^{2}}+\sum_{i=1}^{n_{SW}}a_{1,SW,\log\sigma^{2},i}\times\widetilde{PC}_{i,SW}}{\left(\sum_{i=1}^{n_{SW}}\frac{a_{1,SW,\mu,i}^{2}}{\sigma\left(PC_{i,LW}\right)^{2}}\right)^{1/4}}\right)\\
 & +b
\end{aligned}
\end{equation}

We notice that parts of Equation \ref{fulleq:logvar_expand1} are constant, which means that we can simplify this equation further by defining variance prefactors for longwave radiation ($c_{LW}$; Eq. \ref{constant:c_lw}) and shortwave radiation ($c_{SW}$; Eq. \ref{constant:c_sw}). The standard deviation terms in the normal distributions can now be written in a simpler format:
\begin{equation} \label{logvarterm:LW}
\frac{\exp\left(b_{1,LW,\log\sigma^{2}}+\sum_{i=1}^{n_{LW}}a_{1,LW,\log\sigma^{2},i}\times\widetilde{PC}_{i,LW}\right)}{\left(\sum_{i=1}^{n_{LW}}\frac{a_{1,LW,\mu,i}^{2}}{\sigma\left(PC_{i,LW}\right)^{2}}\right)^{1/4}} = c_{LW}\exp\left(\underbrace{\sqrt{\sum_{i=1}^{n_{LW}}\frac{a_{1,LW,\log\sigma^{2},i}^{2}}{\sigma\left(PC_{i,LW}\right)^{2}}}}_{d_{LW}}\times\left\langle LW\ |\ \Pi_{\log\sigma_{LW}^{2}}\right\rangle_{LW} \right),
\end{equation}

\begin{equation} \label{logvarterm:SW}
\frac{\exp\left(b_{1,SW,\log\sigma^{2}}+\sum_{i=1}^{n_{SW}}a_{1,LW,\log\sigma^{2},i}\times\widetilde{PC}_{i,SW}\right)}{\left(\sum_{i=1}^{n_{SW}}\frac{a_{1,SW,\mu,i}^{2}}{\sigma\left(PC_{i,SW}\right)^{2}}\right)^{1/4}} = c_{SW}\exp\left(\underbrace{\sqrt{\sum_{i=1}^{n_{SW}}\frac{a_{1,SW,\log\sigma^{2},i}^{2}}{\sigma\left(PC_{i,SW}\right)^{2}}}}_{d_{LW}}\times\left\langle SW\ |\ \Pi_{\log\sigma_{SW}^{2}}\right\rangle_{SW} \right).
\end{equation}

Using the same reasoning as for the mean data-driven structure, we find the proportionality coefficient ($\nu$) for the PC loadings of the logarithmic variance structure:
\begin{equation} \label{equivalence:LOGVAR_coeff}
PC_{i,\Pi\log\sigma_{LW}^{2}}=\nu\frac{a_{1,LW,\log\sigma^{2},i}}{\sigma\left(PC_{i,LW}\right)}\ \Rightarrow\ \nu=\left(\sum_{i=1}^{n_{LW}}\frac{a_{1,LW,\log\sigma^{2},i}^{2}}{\sigma\left(PC_{i,LW}\right)^{2}}\right)^{-1/2}.
\end{equation}

We can now express the logarithmic variance structure's PC loadings as:
\begin{equation} \label{logvarterm:PCloading}
PC_{\Pi\log\sigma_{LW}^{2}}=\left(\sum_{i=1}^{n_{LW}}\frac{a_{1,LW,\log\sigma^{2},i}^{2}}{\sigma\left(PC_{i,LW}\right)^{2}}\right)^{-1/2}\frac{a_{1,LW,\log\sigma^{2},i}}{\sigma\left(PC_{i,LW}\right)}.
\end{equation}

For constistency with the mean structures, we transform the logarithmic variance term (Eq. \ref{logvarterm:LW}-\ref{logvarterm:SW}) using the inner product definition:
\begin{equation} \label{logvarterm:gettingD}
c_{LW}\exp\left(\sum_{i=1}^{n_{LW}}\frac{a_{1,LW,\log\sigma^{2},i}\times PC_{i,LW}}{\sigma\left(PC_{i,LW}\right)}\right)=c_{LW}\exp\left(\underbrace{\sqrt{\sum_{i=1}^{n_{LW}}\frac{a_{1,LW,\log\sigma^{2},i}^{2}}{\sigma\left(PC_{i,LW}\right)^{2}}}}_{d_{LW}}\times\left\langle LW\ |\ \Pi_{\log\sigma_{LW}^{2}}\right\rangle_{LW} \right)
\end{equation}

Using this simplification, we derive Equation~\ref{fulleq:goal} in the SI, which is equivalent to Equation 1 in the main text. We will now elaborate upon the short description in the main text on several important aspects of the model training procedure. Specifically, we discuss the cross-validation strategy, hyperparameter settings, and the overall workflow used to train the VED models. 

\subsection{Cross-validation strategy}
We train multiple models on different cross-validation splits to evaluate the sensitivity of model prediction skills to what ensemble members or sensitivity experiments are included in the training data and ensure the trained models are generalizable to out-of-sample data. Since ensemble simulations and sensitivity experiments can be treated as different realizations of the same physical system responding to slightly different forcing, we opt for a data-splitting strategy based on ensemble member labels. 

For Haiyan, 80$\%$ of the ensemble outputs (16 experiments) are used for training, and 20$\%$ of the data (4 experiments) is left for validation and testing. We create different splits by first choosing two experiments as an independent test set; the remaining 18 experiments are partitioned into training and validation subsets by randomly generating a list of two numbers between 1 and 20; the two numbers are then used as references to separate the validation set from the training set. This data-splitting procedure is repeated forty times to fully sample the model variability associated with the choice of data split. To avoid information leakage to the trained models, the test set is truly independent in that the two experiments in it are never used in the training or validation set in all 40 splits. The reference list to create and produce our dataset is shown in Table \ref{table:datasplit}. 

For Maria, we slightly altered the data-splitting strategy due to a lack of samples. The Control (CTRL) simulation is always included in the training set because it has the most samples and represents how the TC evolves in realistic conditions where CRF always exists. The NCRF36 experiment is used as the test dataset amongst the four remaining sensitivity experiments because the storm intensity changes in that experiment depart most from the CTRL simulation. The other experiments are randomly split into a portion that would be merged into the training set (2 experiments) and the other portion for model validation (1 experiment). The cross-validation strategy for Maria yields three different data splits to test the ability of the trained ML models to depict the counterfactual of TC evolution without cloud radiative feedback (CRF).

\subsection{Hyperparameter settings and model optimization}
Next, we describe the hyperparameter settings and how they are tuned to create one probabilistic ML model. All models presented herein are implemented and trained with the PyTorch deep learning library \cite{paszke2019pytorch}. The Adam Optimizer is used to perform stochastic gradient descent for all trained models. The Optuna hyperparameter tuning framework \cite{akiba2019optuna} is used to find the best set of hyperparameters for one particular data split. In the VED framework, there are two main hyperparameters to optimize - the learning rate, which determines the step at which the SGD optimization process proceeds, and the VED loss coefficient. The VED architecture requires a specialized loss function for optimization. The loss function used in this study is defined as follows:

\begin{equation} 
\label{crossval:VEDloss}
\mathrm{VED loss} = \nu \mathrm{Reconstruction\ Loss} + (1-\nu) \mathrm{KL loss},
\end{equation}

where $\nu\in \left[0,1\right]$ is the weight of the reconstruction loss in the overall VED loss. The role of $\nu$ will be expanded upon in the subsequent paragraphs. The objective of the encoder module in the VED architecture can be formulated as finding the posterior probability of latent representation Z given input X ($p(Z|X)$). From Bayes' theorem:
\begin{equation} \label{crossval:bayes}
p(Z|X) = \frac{p(X|Z)p(Z)}{p(X)}
\end{equation}


This posterior is difficult to calculate due to the intractable $p(X)$ term. The VED model solves this problem by having a parameterized prior. We want to make the learned data distribution as close to the prior as possible for regularization. The VED loss implemented in the model can be treated as combining the usual loss function objective for the decoding regression module (the ``reconstruction loss'' in Eq. \ref{crossval:VEDloss}; MAE in this case) and an extra KL regularization term that forces the latent data representation distributions to be close to the prior. The KL regularization term accomplishes the objective of approximating the true conditional probability term $p(X|Z)$ using the predefined prior. Mathematically, this is equivalent to minimizing the Kullback-Leibler (KL) divergence between the prior and the true conditional probability term $p(X|Z)$. By using the ``reparameterization trick'' \citep{kingma2013auto}, the KL loss implemented in this model can be expressed as:
\begin{equation} 
\label{crossval:KLloss}
    \mathrm{KL Loss} = \frac{1}{2} \times \frac{1}{N} \sum_{j=1}^{N} \sum_{k=1}^{K} \left[-1 - \ln\sigma_{jk}^2 + \mu_{jk}^2 + \sigma_{jk}^2\right],
\end{equation}
where N equals batch size, and K represents the width of the latent space. The mean and variance of the learned Gaussian distribution are represented by $\mu$ and $\sigma$. We calculate the $\mu$ and $\sigma$ for every batch ($j$) and latent dimension ($k$) of the training data. Minimizing the summation of these terms ensures that the learned data distribution is close to the prior after training.

For every cross-validation fold, a set of best hyperparameters minimizing training VAE loss is saved. The main hyperparameter to tune in our model is the learning rate. For a given hyperparameter set, seven trials are conducted to account for stochasticity associated with weight initialization. The KL term can result in the model finding a latent representation too close to the uninformative Gaussian prior - a phenomenon known as posterior collapse \cite{Alemi_etal2018}. A KL annealing approach is used to optimize the VED model to avoid this phenomenon. A summary of the KL annealing approach is to train the VED model without KL loss for a certain time and focus on optimizing the decoder prediction task. We continue to train the partially trained model with different strengths of the KL loss by making the $\nu$ coefficient less than 1. We tune $\nu$ manually and save all models with different $\nu$ instead of optimizing it automatically with Optuna because we wanted to show the full sensitivity of model performance to $\nu$. When initializing the optimizer, all parameters are set to default values in PyTorch, with the only exception being the learning rate. Other optimizers are tested, but they did not perform satisfactorily on the prediction task. 280 VED models were trained and stored for the Haiyan ensemble experiments, whereas 21 models were trained for the Maria experiments.

\subsection{Evaluating the trained probabilistic models}
The trained VED models are evaluated with two criteria - good mean prediction skills and a well-calibrated uncertainty in the model outputs. We sample the model spread by running each model 30 times and comparing the 30 model predictions. A suite of stochastic and determinative performance metrics are used to assess the quality of the model. Two stochastic metrics are evaluated: the CRPS score and the SSREL value ~\cite{Haynes_etal2023}. 

The CRPS score compares the Cumulative Distribution Function (CDF) of the probabilistic forecasts against the observations; it is also a generalization of the deterministic mean absolute error (MAE) metric for probabilistic models:
\begin{equation} \label{crossval:CRPS}
\mathrm{CRPS}\left(F, y_{true}\right) = \int_{-\infty}^{\infty} \left[F\left(y_{pred}\right)-\mathcal{H}(y_{pred}-y_{true})\right]^2 dy_{pred},
\end{equation}
where $F$ represents the CDF of the model predictions, $\mathcal{H}$ is the Heaviside step function applied to the difference between the truth ($y_{true}$) and one prediction ($y_{pred}$) sampled from the full distribution. By definition, the Heaviside function outputs 1 if the prediction is larger than the truth and 0 in the opposite case. A well-calibrated model should have as small a CRPS score as possible.

The SSREL value \cite{Haynes_etal2023} measures the quality of a binned spread-skill plot. The binned spread-skill plot is an assessment of the statistical consistency of a probabilistic model \cite{DelleMonoche_etal2013}. Statistical consistency refers to the situation where the probabilistic model likely samples from the same underlying distribution as that in the truth. A statistically consistent model should have its spread closely match its error. If the spread-skill curve of a model deviates from the 1-1 line in a binned spread-skill plot, the model is either under-dispersive (overconfident) or over-dispersive (underconfident), both being undesirable traits of a probabilistic prediction model. The SSREL value measures weighted distances between the model curve and the 1-1 line:
\begin{equation} \label{crossval:SSREL}
\mathrm{SSREL} = \sum_{k=1}^{K} \frac{N_k}{N} \left[\mathrm{RMSE}_k-\overline{\mathrm{SD}_{k}}\right],
\end{equation}
where $K$ is the number of bins, $N_{k}$ is the number of samples in a bin, $N$ is the total number of samples, $\mathrm{RMSE}_{k}$ is the root-mean-square-error of the model predictions for samples within the bin, $\mathrm{SD}$ is the standard deviation of the model predictions. 

A perfectly calibrated model will have a model curve that overlaps the 1-1 line, which corresponds to an SSREL value of 0. We report two metrics for the mean deterministic skills: the mean absolute error (MAE) and root mean squared error (RMSE).

\subsection{Model baseline}
\begin{figure}
    \includegraphics[width=\textwidth]{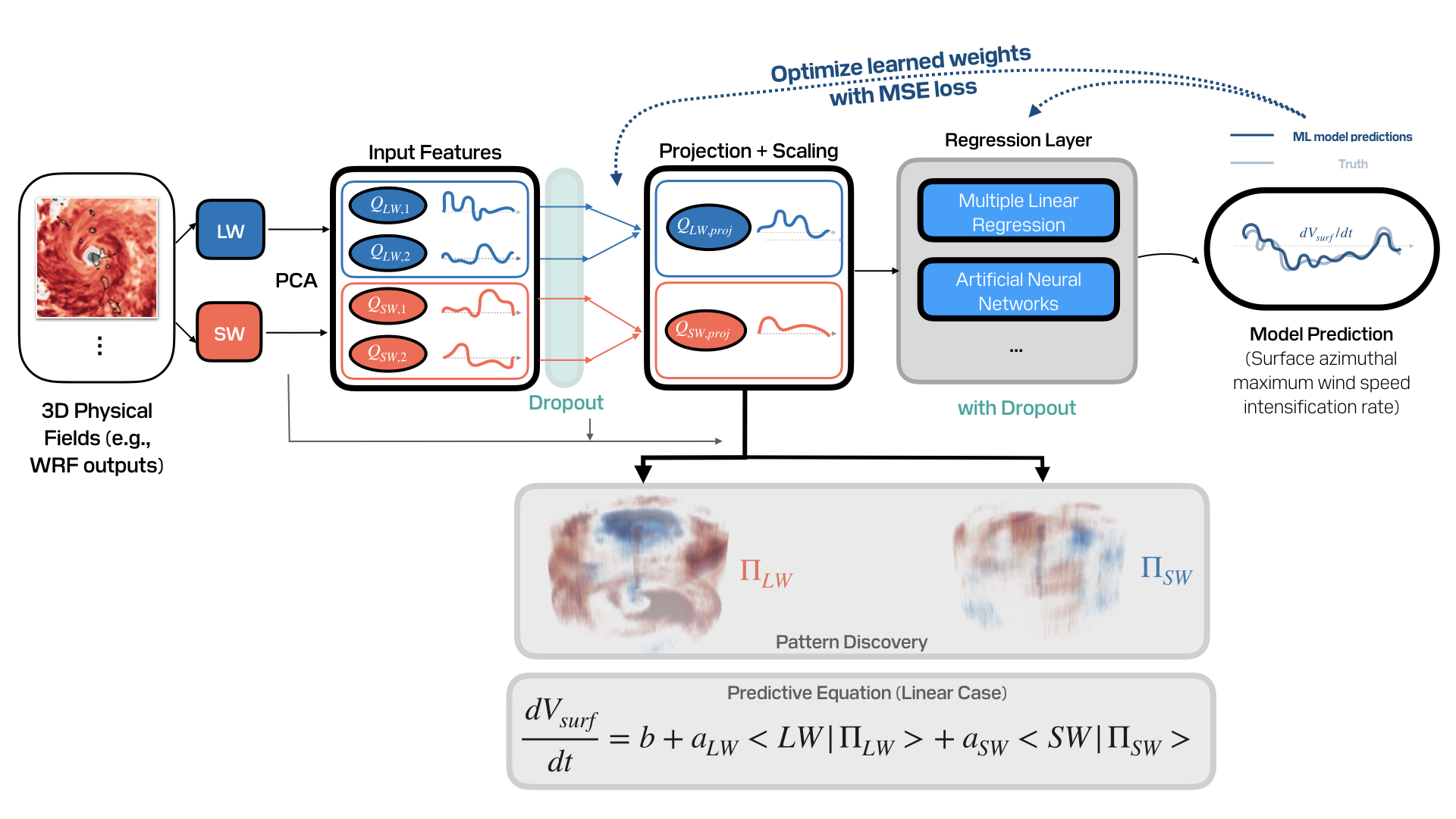}
    \caption{The interpretable linear model baseline to compare to the VED model. This baseline model introduces uncertainty with the dropout operation.}
    \label{fig:SIschematic}
\end{figure}

We implement a simple two-layer linear regression model (Fig. \ref{fig:SIschematic}) as a baseline to examine if the more complex VED architecture improves the overall prediction skills and provides a more well-calibrated prediction uncertainty. This baseline is analogous to a two-branch principal component regression. We opt for optimizing it through stochastic gradient descent to enhance flexibility in selecting the loss function, in this case, Mean Absolute Error (MAE).

Compared to the linear VED case, the baseline model only extracts two structures from the longwave and shortwave radiation information. These structures are analogous to the $\mu_{LW}$ and $\mu_{SW}$ in the VED model. Rather than extracting the uncertainty structures and sampling them with the reparameterization trick, we introduce uncertainty in the baseline model by using a dropout mechanism. The dropout mechanism zeroes out a random selection of input sections, which allows the model to have uncertainties in both the structure layer level and the final prediction outputs. Dropout layers are added to both channels in the first structure extraction layer and the final prediction layer, allowing the quantification of uncertainties in the structures and the predictions. The amount of input features dropped by this operation is determined by the $\mathbf{dropout\ rate}$ hyperparameter, a hyperparameter that ranges from 0 to 1. Larger $\mathbf{dropout\ rate}$ values mean more input features are kept during inference, and vice versa.

\subsection{Choosing the best VED model and comparison with the best baseline model}
\begin{figure}
    \includegraphics[width=\textwidth]{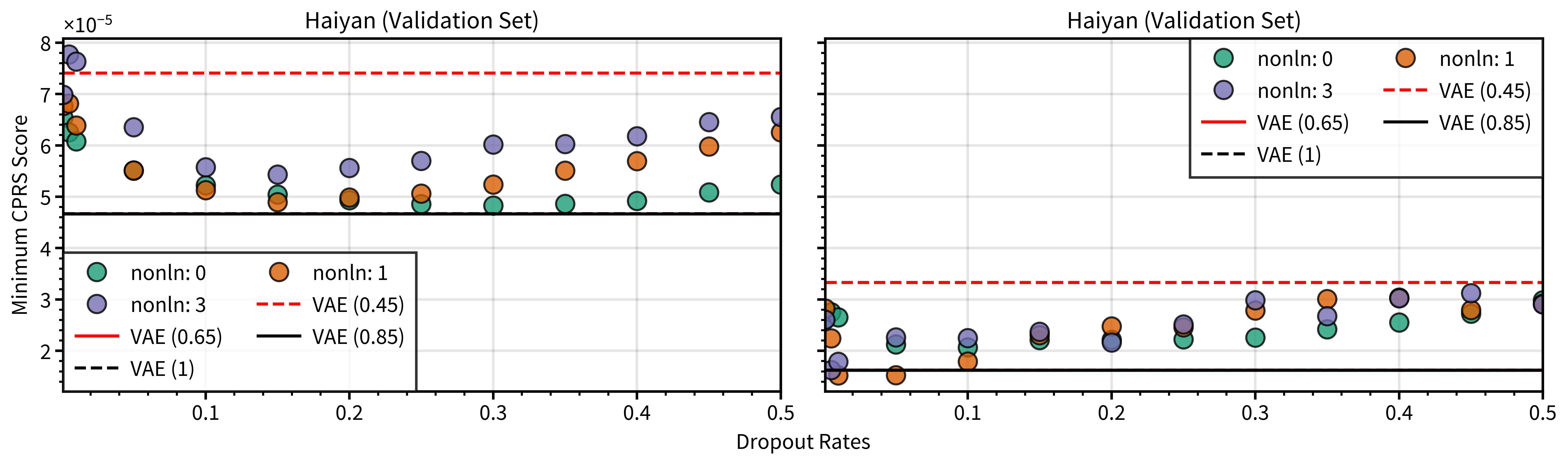}
    \caption{The median (left) and minimum (right) CRPS scores for the baseline and VED models on the validation sets. The effect of nonlinearity on the baseline models is tested by adding 1, 3, or 5 regression layers with Rectified Linear Unit (ReLU) activation functions. Both types of models exhibit an optimal range of hyperparameters that would result in the best skill scores. However, the best VED model overperforms the best baseline model.}
    \label{fig:SI_CRPS_valid}
\end{figure}

Here, we describe the strategy to choose the best model to produce the results shown in the main text. The best model is chosen objectively based on the stochastic metrics discussed in the previous section. Figure \ref{fig:SI_CRPS_valid} shows the minimum and median CRPS scores for all trained Haiyan VED and baseline models with different hyperparameters on the validation set. This figure allows us to compare the models with discern the sensitivity of model skills towards critical hyperparameters related to stochasticity - the $\mathbf{dropout\ rate}$ for the baseline models, and the VED loss coefficient ($\lambda$) for the VED models. We also substitute the fully linear prediction layer in the baseline model with feed-forward neural networks for different depths to evaluate if adding nonlinearity to the models improves the prediction skills.

\begin{figure}
    \includegraphics[width=\textwidth]{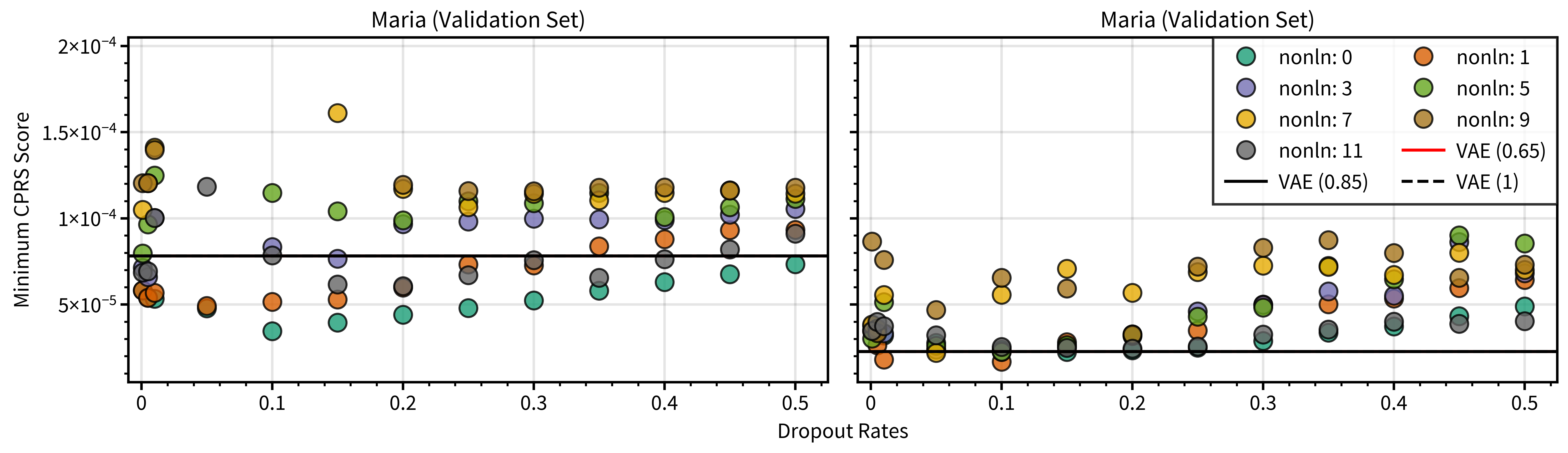}
    \caption{CRPS performance diagram for the Maria validation set. For Maria, we tested the effect of adding nonlinearity up to 11 nonlinear regression layers.}
    \label{fig:SI_CRPS_valid_maria}
\end{figure}

Adding nonlinearity degrades the median CRPS scores for all models for most $\mathbf{dropout\ rates}$ (Fig. \ref{fig:SI_CRPS_valid},\ref{fig:SI_CRPS_valid_maria}), which justifies keeping the model fully linear. The reason why a fully linear model is sufficient here is likely due to the low sample regime situation we have for this particular problem. The median CRPS score curves for all baseline models (Fig. \ref{fig:SI_CRPS_valid},\ref{fig:SI_CRPS_valid_maria}) are parabola-shaped, with distinct minima, suggesting the existence of an optimal range of $\mathbf{dropout\ rates}$ that is desirable for better model generalizability.

We now turn our focus to the VED models. Adding the median VED CRPS scores for all trained VED models with different $\lambda$ shows deteriorated prediction skills for a $\lambda$ that is too small, i.e., too large a KL loss during training. The CRPS values for the minimum VED model with a given $\lambda$ show that the best model is created when $\lambda$ equals 0.85.

The CRPS score comparison above establishes optimal values for the $\mathbf{dropout\ rates}$ (the baseline models) and $\lambda$ (the VED models). The best models for comparison are determined by calculating the SSREL scores for all models trained with the optimal coefficients. 

\begin{figure}
    \includegraphics[width=0.75\textwidth]{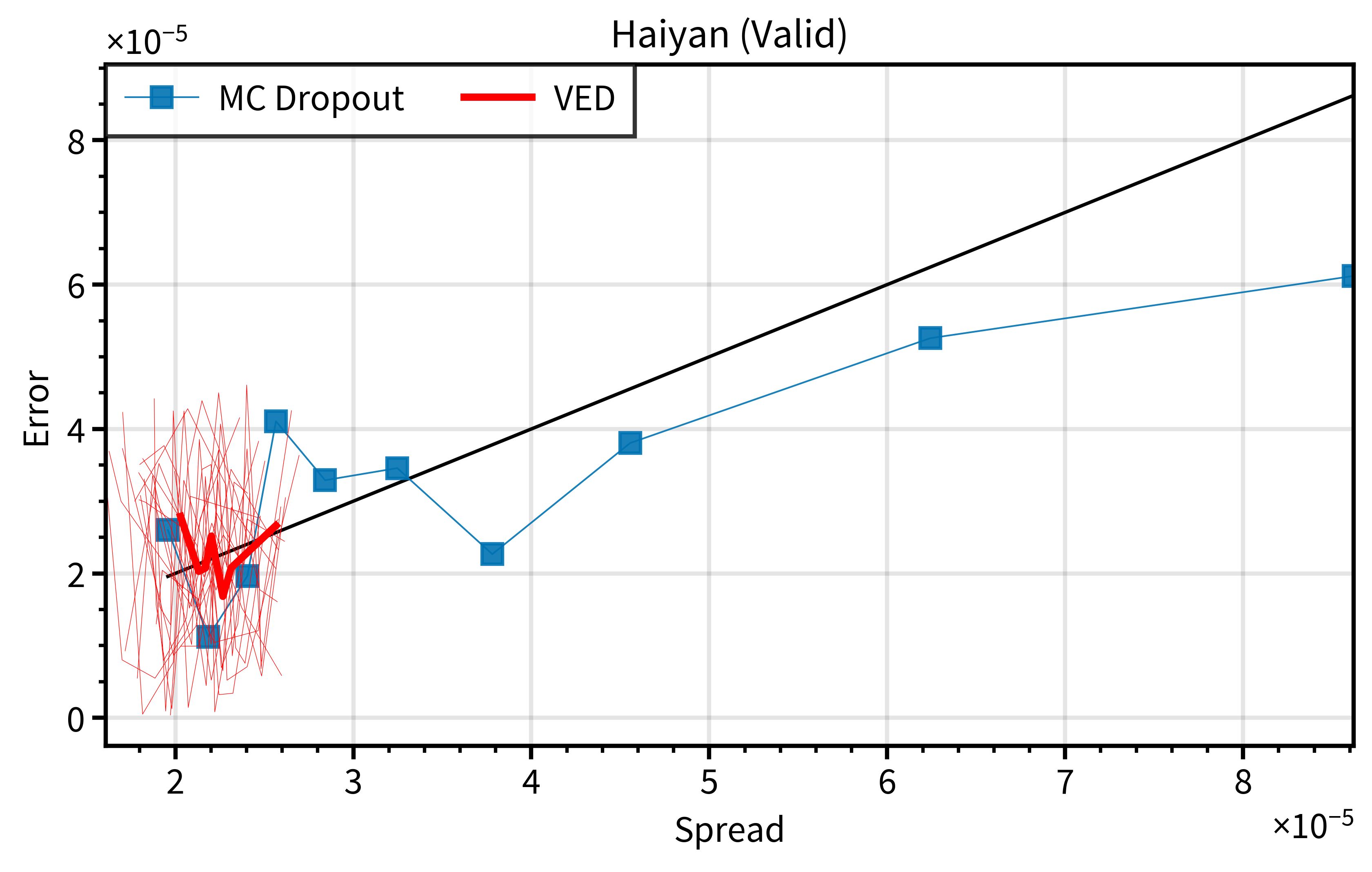}
    \caption{Mean spread-skill curves for the best-performing baseline model (black) and the best-performing VED model (red thick). The VED curves for individual predictions with different random seeds are shown to show the prediction spread captured by the VED model.}
    \label{fig:SI_spreadskill_haiyan}
\end{figure}

The spread-skill plot for the best baseline and VED models (Fig. \ref{fig:SI_spreadskill_haiyan}) provides a strong justification for using the VED model in our study. Specifically, the spread-skill curve of the VED model is much shorter than the baseline one, which indicates the VED predictions are more accurate. Plotting the full predictions of the best VED and baseline model shows that the VED better captures the peaks in the training dataset. The VED model also removes the large biases in early intensification rates seen in the baseline predictions on the test set. The VED curve is much closer to the 1-1 line than the baseline model curve. Based on these comparisons, we conclude that the VED model is superior to the baseline model for our research task because it makes smaller prediction errors and has better-calibrated uncertainty.

\begin{figure}
    \includegraphics[width=0.75\textwidth]{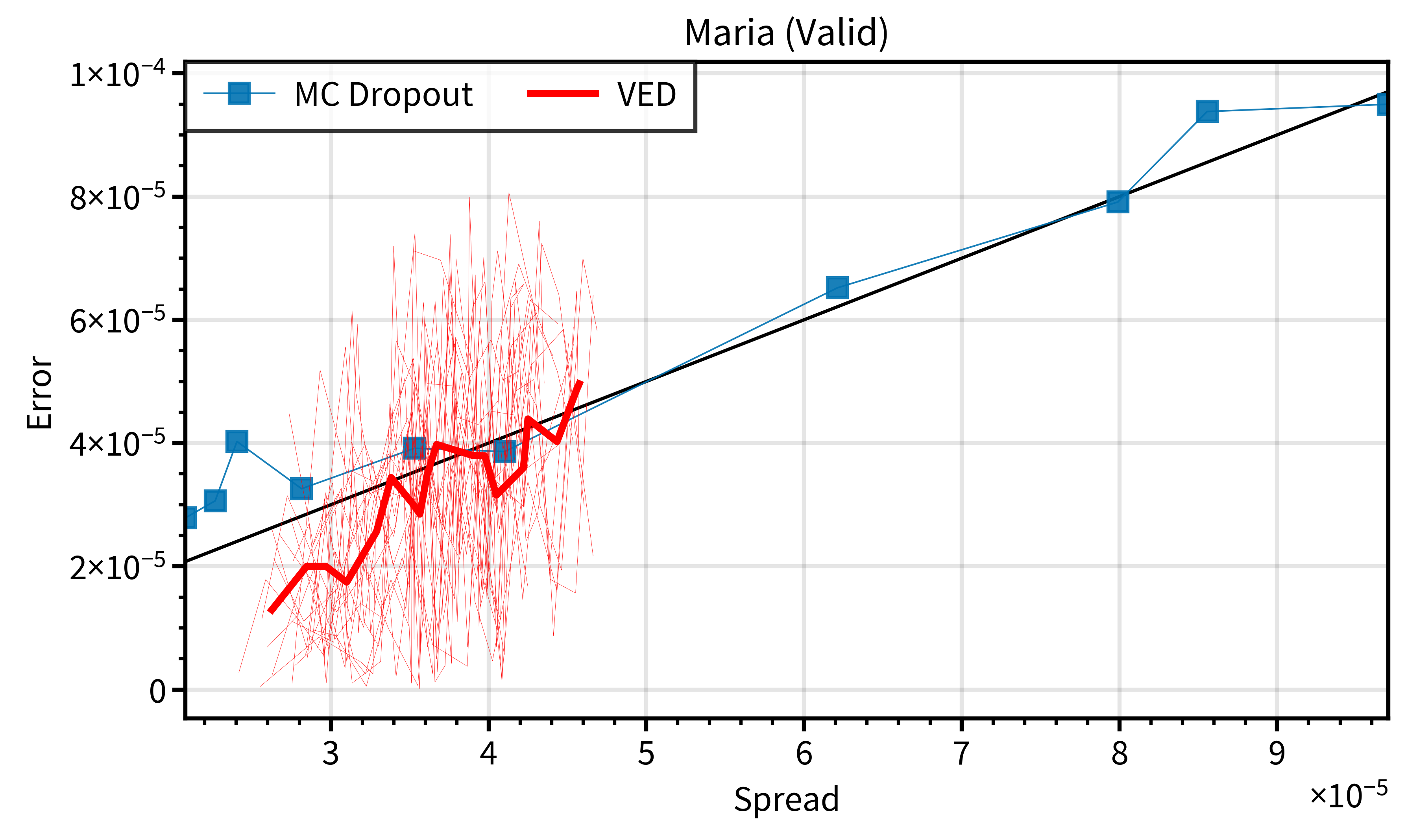}
    \caption{The best-performing VED model for Maria is less accurate than the best Haiyan model on their corresponding validation set. However, the VED model somewhat outperforms the baseline model, especially regarding prediction errors.}
    \label{fig:SI_spreadskill_maria}
\end{figure}

For Maria, the best VED model performs similarly to the best baseline model in terms of the minimum CRPS score (Fig. \ref{fig:SI_CRPS_valid_maria}). The uncertainty for both the best baseline model and best VED model are well calibrated. However, the VED model is again preferable for Maria because of the smaller prediction errors (Fig. \ref{fig:SI_spreadskill_maria}).

Interestingly, the benefit of the VED model compared to the baseline model seems to scale to the sample size. The VED model always overperforms the baseline for the Haiyan ensemble case with a larger sample size (Table \ref{table:SIhaiyan}), whereas the VED model mostly only overperforms the baseline in probabilistic metrics for the Maria simulations (Table \ref{table:SImaria}).

\section{Scientific Interpretation with the Machine Learning Framework}
This section complements Section 4 in the main text and elaborates on the procedure we took to get some key physical interpretations with our machine learning model. Specifically, we rely on the fully linear nature of the ML framework discussed in Section \ref{Details:Maths}. This allows us to design targeted sensitivity analysis to address specific scientific problems.

\subsection{Decomposition by Vertical Levels}
This section provides support for the statement in the main text that ``shallow cloud contribution to intensification is between two times and one order of magnitude smaller than the upper-level radiative contribution''. For this purpose, we discuss the scientific framing, preprocessing of data, and the relevant sensitivity experiment results.

\paragraph{Methodology}
We isolate the effect of heating anomalies in different vertical levels by zeroing out all grid point values in the input data \textit{except grid points at the interested vertical level}. Projecting the perturbed inputs on the PC eigenvectors gives a new set of PC coefficients that the ML model can use. The model predictions with the new PC inputs represent the contribution radiation at the interested vertical level has on the overall intensification forecast. In other words, we frame these sensitivity experiments as a linear decomposition problem.

Taking longwave radiation as an example, we can separate the LW field into three separate terms:
\begin{equation} \label{sensitivity:LWdecomp}
LW = LW_{100} + LW_{1000} + LW_{200-900},
\end{equation}
where $LW_{100}$ ($LW_{1000}$) is the longwave field where all grid points except those at 100 (1000) hPa are zeroed out, $LW_{200-900}$ is the longwave field where grid points at 100 and 1000 hPa are zeroed out.

Projecting the three components of the right hand side of Eq.~\ref{sensitivity:LWdecomp} onto the longwave PC eigenvectors yields the same decomposition for the PC loadings:
\begin{equation} \label{sensitivity:PCdecomp}
\forall i \in \llbracket 1,n_{LW} \rrbracket \ , \  PC_{i,LW} = PC_{i,LW_{100}} + PC_{i,LW_{1000}} + PC_{i,LW_{200-900}}
\end{equation}

After standardization, Equation (\ref{sensitivity:PCdecomp}) can be separated into a constant part: 

\begin{equation} \label{sensitivity:PCstandardization_const}
\underbrace{\left(\frac{\overline{PC_{i,LW_{100}}}+\overline{PC_{i,LW_{1000}}}+\overline{PC_{i,LW_{200-900}}}}{\sigma(PC_{i,LW})}\right)}_{\overline{PC_{i,LW}}}
\end{equation}

and a part that varies in time:
\begin{equation} \label{sensitivity:PCstandardization_PC}
\underbrace{\left[\frac{PC_{i,LW_{100}}}{\sigma(PC_{i,LW})} +  \frac{PC_{i,LW_{1000}}}{\sigma(PC_{i,LW})} + \frac{PC_{i,LW_{900-200}}}{\sigma(PC_{i,LW})}\right]}_{PC_{i,LW}},
\end{equation}

Substituting Equation \ref{sensitivity:PCstandardization_PC} into Equation \ref{fulleq:a2_inside_N} and expanding the first term in Equation \ref{fulleq:a2_inside_N}, we can decompose surface intensification into additive terms that match the longwave radiative heating field's decomposition of Eq. \ref{sensitivity:LWdecomp}.

\begin{equation} \label{sensitivity:define}
\begin{aligned}
\frac{dV_{surf, 24}}{dt} = &\left(\frac{dV_{LW, 100 hPa}}{dt}+\frac{dV_{LW, 200-1000 hPa}}{dt}\right) \\
&+ \frac{dV_{SW, 100-1000 hPa}}{dt}+b,
\end{aligned}
\end{equation}
where $\frac{dV_{LW, 100 hPa}}{dt}$ quantifies how radiation anomalies at 100 hPa affect the surface intensification.

\begin{figure}
    \includegraphics[width=\textwidth]{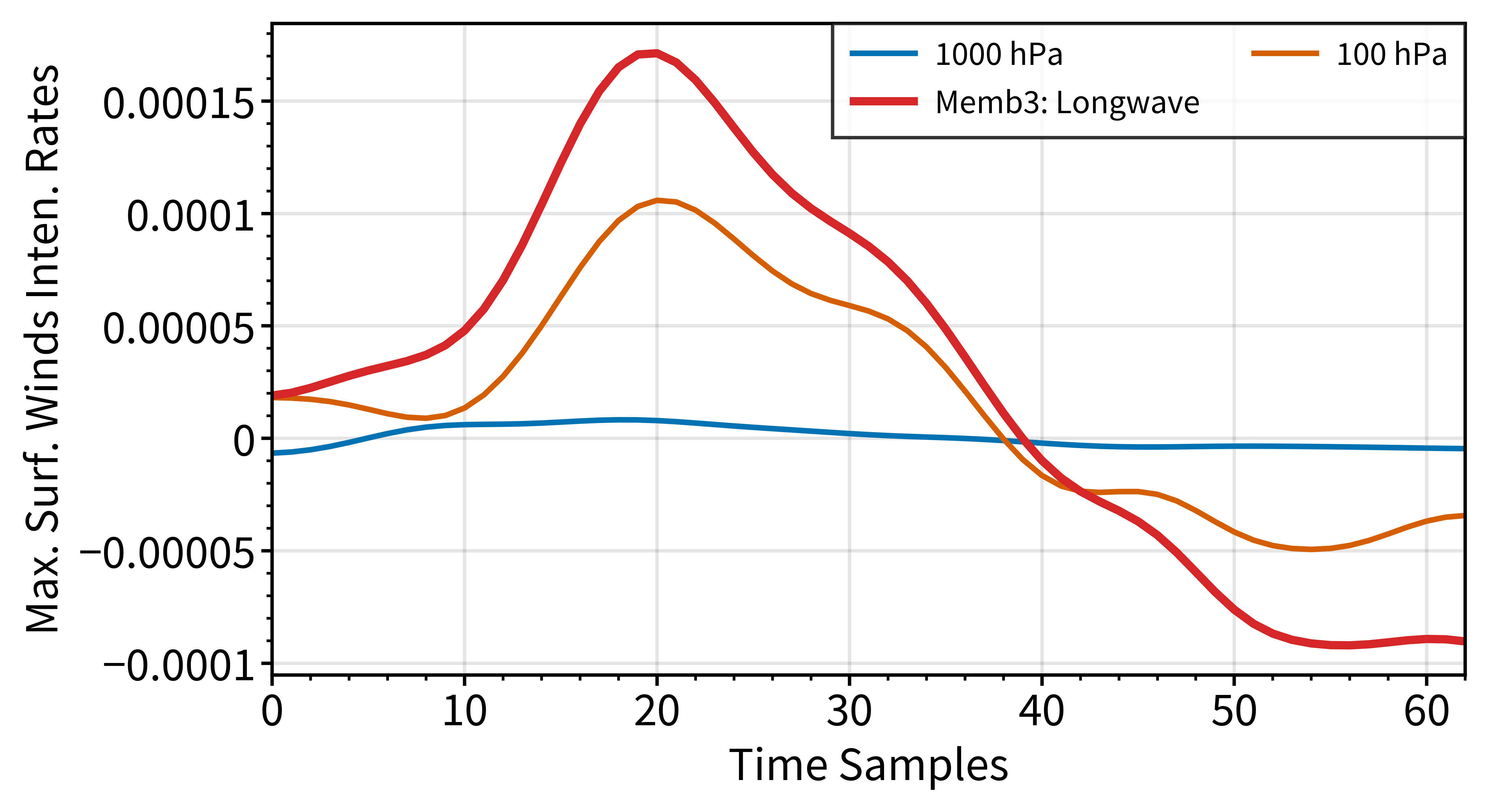}
    \caption{Linear decomposition of the longwave contribution to the intensification of Haiyan. The contribution of upper-level radiative anomalies is consistently larger than the contribution of the low-level radiative anomalies.}
    \label{fig:SI_decompose_haiyan}
\end{figure}

\paragraph{Result}
A key interpretation of this paper relies on identifying a suitable vertical level where radiation likely yields the most effect on TC genesis. Decomposing the predictions into different components allows us to quantify their relative impacts. Here, we define radiation anomaly at 100 hPa as a proxy for the deep convection radiation effect and 1000 hPa as a proxy for the shallow cloud radiation. Figure \ref{fig:SI_decompose_haiyan}, which compares the decomposed longwave contributions at two vertical levels, shows the intensification associated with upper-level radiation to consistently be larger than the intensification associated with the radiative anomalies at 1000 hPa. This is true across all Haiyan ensemble members and almost all periods during the TC life cycle. This result leads us to focus mostly on the upper-level radiative anomalies when we interpret the extracted structures. Note that our model's structure does not allow us to find a direct causal link between the shallow cloud radiative feedback, deep convective cloud radiation, and surface wind intensification.

\section{Frozen Moist Static Energy Variance Budget}
The Frozen Moist Static Energy Spatial Variance Budget used in the main text follows the approach of \cite{WingEEmanuel2014SelfAG}. Frozen Moist Static Energy ($h$), a variable conserved during moist adiabatic processes, can be expressed as:

\begin{equation} \label{eq:MSE_definition}
h = C_p T + gz + L_v q_v - L_f q_{ice},
\end{equation}

where $L_{v}$ is the latent heat of evaporation, $L_{f}$ is the latent heat of fusion, $q_v$ is the mixing ratio of water vapor, $q_{ice}$ is the mixing ratio of ice condensates, and $T$ is the temperature.

The vertically integrated $h$ ($\hat{h}$) has appealing properties, including that it remains conserved during convective transport. The only sources for $\hat{h}$ in the atmosphere are net column longwave radiative flux convergence ($\mathrm{NetLW}$), net column shortwave radiative flux convergence ($\mathrm{NetSW}$), latent heat flux ($\mathrm{LHF}$), and sensitive heat flux ($\mathrm{SHF}$). The only sink of $\hat{h}$ is the horizontal divergence of $\hat{h}$ flux.

In other words, the MSE budget equation is,
\begin{equation} \label{eq:MSE_budget}
\frac{\partial \hat{h}}{\partial t} = LHF + SHF + NetLW + NetSW - \mathbf{\nabla}_h \cdot \vec{u} \hat{h},
\end{equation}

Subtracting the domain mean from Equation \ref{eq:MSE_budget} results in a budget equation for the time evolution of the anomalies in vertically integrated MSE,
\begin{equation} \label{eq:MSE_anomaly_budget}
\frac{\partial \hat{h}^\prime}{\partial t} = LHF^\prime + SHF^\prime + NetLW^\prime + NetSW^\prime - \mathbf{\nabla}_h \cdot \vec{u} \hat{h},
\end{equation}
where $\prime$ represents the anomaly of each field with respect to the domain mean.

The MSE variance budget equation can be obtained by multiplying Equation \ref{eq:MSE_anomaly_budget} with $\hat{h}^\prime$,
\begin{equation} \label{eq:MSE_spatialvariance_budget}
\frac{1}{2} \frac{\partial \hat{h}^{\prime^2}}{\partial t} = \underbrace{\hat{h}^{\prime} \mathrm{LHF}^\prime + \hat{h}^{\prime} \mathrm{SHF}^\prime}_{\mathrm{SEF Contribution}} + \underbrace{\hat{h}^{\prime} \mathrm{NetLW}^\prime + \hat{h}^{\prime} \mathrm{NetSW}^\prime}_{\mathrm{Radiative Contribution}} - \mathbf{\nabla}_h \cdot \vec{u} \hat{h},
\end{equation}

\noindent where $\mathrm{SEF}$ (the Surface Enthalpy Flux) is the sum of $\mathrm{LHF}$ (the Latent Heat Flux) and $\mathrm{SHF}$ (the Sensible Heat Flux). The radiative contribution consists of net column longwave radiative flux convergence ($\mathrm{NetLW}$) and net column shortwave radiative flux convergence ($\mathrm{NetSW}$). Primes are used to indicate anomalies relative to the mean of the spatial domain, which is represented by overlines. MSE spatial variance source terms are obtained by spatially averaging all terms on the right-hand side of Equation ~\ref{eq:MSE_spatialvariance_budget}.

\end{article}
\clearpage

\bibliography{agusample}


\begin{table}
\centering
\caption{Prediction skills of the best VED and the best baseline model on the intensification of Haiyan. Also shown in the table are the median skills of all trained models (numbers in brackets). All values in the table are multiplied by $10^5$ for visualization purposes. The best model is illustrated with bolded numbers.}
\begin{tabular}{l|l|l|l|l}
Experiment                & Metric & Training    & Validation & Test \\ \hline
\multirow{4}{*}{VED}      & CRPS   & \textbf{2.83} (3.35) &  \textbf{1.65} (4.69)          &  \textbf{2.86} (3.73)    \\
                          & SSREL  & \textbf{1.31} (2.61) &  \textbf{0.33} (3.54) &  \textbf{1.03} (2.44)    \\
                          & RMSE   & \textbf{4.52} (5.34) &  \textbf{2.61} (6.99) &  \textbf{4.83} (5.95)    \\
                          & MAE    &  \textbf{3.67} (4.25) &  \textbf{2.15} (5.76) & \textbf{3.75} (4.74)     \\ \cline{1-1}
\multirow{4}{*}{Baseline} & CRPS   &  3.23 (3.48) &  2.26 (4.83) & 3.45 (3.90)     \\
                          & SSREL  &  1.89 (2.45) &  0.67 (4.19) & 1.17 (2.76) \\
                          & RMSE   &  5.35 (5.68) &  3.67 (7.17) & 5.54 (6.09) \\
                          & MAE    &  4.28 (4.56) &  2.96 (6.00) & 4.31 (4.93)  \\ \hline
\end{tabular}
\label{table:SIhaiyan}
\end{table}

\begin{table}
\centering
\caption{Prediction skills of the best VED and the best baseline model on the intensification of Maria.}
\begin{tabular}{l|l|l|l|l}
Experiment                & Metric & Training    & Validation & Test \\ \hline
\multirow{4}{*}{VED}      & CRPS   & 1.76 (5.88) &  \textbf{2.28} (7.78)          &  \textbf{1.39} (5.26) \\
                          & SSREL  & 0.59 (3.51) & \textbf{0.80} (4.33) &  \textbf{0.33} (2.95) \\
                          & RMSE   & 3.74 (9.27) &  3.93 (10.12) &  \textbf{2.21} (7.32) \\
                          & MAE    &  2.21 (7.95) & 3.20 (9.95) & \textbf{1.74} (7.08) \\ \cline{1-1}
\multirow{4}{*}{Baseline} & CRPS   &  \textbf{1.16} (2.32) &  2.39 (3.46) & 2.32 (2.60) \\
                          & SSREL  &  \textbf{0.36} (1.62) &  0.88 (3.28) & 1.68 (2.21) \\
                          & RMSE   &  \textbf{2.46} (4.57) &  \textbf{3.57} (6.22) & 3.54 (3.84) \\
                          & MAE    &  \textbf{2.03} (3.21) &  \textbf{3.13} (4.85) & 3.22 (3.51)  \\ \hline
\end{tabular}
\label{table:SImaria}
\end{table}

\begin{table}
\centering
\caption{Characteristics of the data splits used for training the Haiyan models. Ensemble Members that are not numbered in the table are treated as training sets.}
\begin{tabular}{l|ll|ll|l|ll|ll}
Data split & \multicolumn{2}{l|}{Validation} & \multicolumn{2}{l|}{Test} & Data split & \multicolumn{2}{l|}{Validation} & \multicolumn{2}{l}{Test} \\
1          & 11              & 4             & 10          & 17          &  20        & 16              &  11           & 10          & 17            \\
2          & 11              & 7             & 10          & 17          &  21        & 16              &  12           & 10          & 17           \\
3          & 12              & 13            & 10          & 17          &  22       & 18                & 11              & 10          & 17         \\
4          & 12              & 14            & 10          & 17         &   23       & 18                &  1             & 10          & 17           \\
5          & 12              & 6             & 10          & 17          &  24       & 18                &  3            & 10          & 17           \\
6          & 13              & 19            & 10          & 17           & 25        & 19                &  1             & 10          & 17         \\
7          & 13              & 7             & 10          & 17           & 26        & 19                &  7             & 10          & 17         \\
8          & 14              & 12            & 10          & 17           & 27        & 19                &  8             & 10          & 17          \\
9          & 14              & 18            & 10          & 17          &  28        & 1                &   11            & 10          & 17           \\
10         & 14              & 19            & 10          & 17          &  29        & 1                &   5            & 10          & 17            \\
11         & 14              & 4             & 10          & 17          &  30        & 1               &   9           & 10          & 17           \\
12         & 14              & 6             & 10          & 17          &  31        & 2               &   19           & 10          & 17           \\
13         & 14              & 7             & 10          & 17          &  32        & 2                &  7             & 10          & 17            \\
14         & 15              & 13            & 10          & 17           & 33        & 3                &  19            & 10          & 17         \\
15         & 15              & 16            & 10          & 17          &  34       & 3                & 1              & 10          & 17           \\
16         & 15              & 19            & 10          & 17          &  35      &  4               & 6              & 10          & 17            \\
17         & 15              &  5            & 10          & 17          &  36       & 5                & 19              & 10          & 17          \\
18         & 15              &  7            & 10          & 17          &  37       & 6                &  1             & 10          & 17         \\
19        & 15              &  8            & 10          & 17         &    38      & 7                &  14             & 10          & 17         \\
39        & 7                &  3             & 10          & 17       &    40      & 7                &  5             & 10          & 17          \\
\end{tabular}
\label{table:datasplit}
\end{table}
